\documentclass[a4paper,fleqn,usenatbib]{mnras}

\usepackage[T1]{fontenc}
\usepackage{ae,aecompl}

\usepackage{graphicx}	
\usepackage{amsmath}	
\usepackage{amssymb}	

\title[Non-gravitational feedback in galaxy clusters] 
{Excess entropy and energy feedback from within cluster cores up to $r_{200}$}
\author[Iqbal et~al.]
{Asif Iqbal$^{1}$\thanks{
asifiqbal@kashmiruniversity.net}, Subhabrata Majumdar$^{2}$\thanks{
subha@tifr.res.in}, Biman B. Nath$^{3}$\thanks{
biman@rri.res.in}, Stefano Ettori$^{4,5}$\thanks{stefano.ettori@oabo.inaf.it}, 
\newauthor Dominique Eckert$^{6}$\thanks{Dominique.Eckert@unige.ch} and Manzoor A. Malik$^{1}$\thanks{
mmalik@kashmiruniversity.ac.in}\\
  $^{1}$Department of Physics, University of Kashmir, Hazratbal, Srinagar, J\&K, 190011, India \\
  $^{2}$Tata Institute of Fundamental Research, 1 Homi Bhabha Road, Mumbai, 400005, India\\
  $^{3}$Raman Research Institute, Sadashiva Nagar, Bangalore, 560080, India\\
  $^{4}$INAF, Osservatorio Astronomico di Bologna, Via Ranzani 1, I-40127, Bologna, Italy \\
  $^{5}$INFN, Sezione di Bologna, viale Berti Pichat 6/2, 40127, Bologna, Italy\\
  $^{6}$Astronomy Department, University of Geneva 16, ch. d'Ecogia, CH-1290 Versoix, Switzerland
} 

\date{}

\begin{document}

\maketitle
 
\begin{abstract}
We estimate the ``non-gravitational'' entropy-injection profiles, $\Delta K$, and the resultant energy feedback profiles, $\Delta E$, of the intracluster medium for 17 clusters using their Planck SZ  and ROSAT X-Ray observations, spanning a large radial range from $0.2r_{500}$ up to $r_{200}$.  The feedback profiles are estimated by comparing the observed entropy, at fixed gas mass shells, with theoretical entropy profiles predicted from non-radiative hydrodynamic simulations. We include non-thermal pressure and gas clumping in our analysis. The inclusion of non-thermal pressure and clumping  results in changing the estimates for $r_{500}$ and $r_{200}$ by 10\%-20\%. When clumpiness is not considered it leads to an under-estimation of $\Delta K\approx300$ keV cm$^2$ at $r_{500}$ and $\Delta K\approx1100$ keV cm$^2$ at $r_{200}$. On the other hand, neglecting non-thermal pressure results in an over-estimation of $\Delta K\approx 100$ keV cm$^2$ at $r_{500}$ and under-estimation of $\Delta K\approx450$ keV cm$^2$ at $r_{200}$. For the estimated feedback energy, we find that ignoring clumping leads to an under-estimation of  energy per particle $\Delta E\approx1$ keV at $r_{500}$ and $\Delta E\approx1.5$ keV at $r_{200}$. Similarly, neglect of the non-thermal pressure results in an over-estimation of  $\Delta E\approx0.5$ keV at $r_{500}$ and under-estimation of $\Delta E\approx0.25$ keV  at $r_{200}$.  We find entropy floor of $\Delta K\approx300$ keV cm$^2$ is ruled out at $\approx3\sigma$ throughout the entire radial range and  $\Delta E\approx1$ keV at more than 3$\sigma$ beyond $r_{500}$, strongly constraining ICM pre-heating scenarios.  We also demonstrate robustness of results w.r.t sample selection, X-Ray analysis procedures, entropy modeling etc. 
\end{abstract}
\section{INTRODUCTION}
Clusters of galaxies are the largest evolved structures in the universe and, as such,
qualify for being important cosmological probes. The abundance of galaxy clusters provides sensitive constraints on the cosmological parameters that govern the growth of structures in the universe \citep{0,1,2}. In this regard, X-ray observations provide a useful tool for identifying 
and studying galaxy clusters \citep{13,10,Eckert2013a,Eckert2013b}. Galaxy clusters can also be observed in the microwave band through Sunyaev-Zel'dovich (SZ) effect \citep{Sunyaev1972,Sunyaev1980}, which results from the up-scatting of cosmic microwave background (CMB) photons by hot electrons in the intracluster
medium (ICM). The SZ effect has a unique property that unlike X-ray emission it is independent of redshift and does not suffer from cosmological dimming. With the current and upcoming data 
from Planck, the SZ cluster surveys have  become a robust probe for determining cosmological parameters and global properties of ICM \citep{Planck2013a,Planck2014,Eckert2013a,McCarthy2014,Ettori2015}.

However, in order to obtain robust cosmological estimates using such techniques one requires the precise knowledge of the evolution of galaxy clusters with redshift and the thermodynamical properties of ICM. 
In the simplest case, where one considers a pure gravitational collapse, the cluster scaling relations are expected to follow simple self-similarity \citep{Kaiser1986}. 
Correlations between the X-ray properties are widely used to probe the self-similarity in the galaxy clusters.
For example, the luminosity-temperature ($L_x-T$) relation for self-similar models predict a
shallower slope ($L_x\propto T^{~2}$) than observed ($L_x\propto T^{~3}$) \citep{Edge1991,Markevitch1998}  implying a break in the self-similarity in galaxy clusters. 
Similarly, studies of the scaling relations involving SZ effect also show discrepancies between observations and predictions from a pure gravitational model \citep{Holderb2001,Battaglia2012}. 
Such studies have revealed the importance of the  complex non-gravitational processes, such as injection of energy feedback from active galactic nuclei, radiative cooling, supernovae, and star formation, influencing the thermal structure of ICM, particularly in low mass (temperature) clusters \citep{Voit2005,Roychowdhury2005,Chaudhuri2012,Chaudhuri2013}.

The first direct evidence for non-gravitational entropy in galaxy clusters and galaxy groups was given by \cite{David1996} using  ROSAT Position Sensitive Proportional Counter (PSPC) observations.
 The observations showed that there was excess entropy compared to that expected from gas collapsing in a gravitational potential.
Motivated by these findings, several groups have drawn similar conclusions using both numerical and semi analytical models  with an entropy floor of the order of $300-400$ keV cm$^2$ \citep{Ponman1999,Tozzi2001,Eckert2013a,Chaudhuri2012,Chaudhuri2013}.
Although, SNe feedback are essential to explain the enrichment of the ICM to the observed metallicity level and heavy-element abundances, they provide insufficient amount of energy per particle as compared 
to recent observations. Moreover, they are also inefficient to quench cooling in massive galaxies \citep{Springel2005}.  
There is a growing evidence that AGN feedback mechanism provides a major source of heating for the ICM gas, thereby reducing number of cooling flow clusters 
\citep{McNamara2007,Gaspari2011,Chaudhuri2012,Chaudhuri2013,Gaspari2014}. The  AGN-jet simulations show that such mechanisms can overcome the cooling flow over a cosmological timescale and produce results similar to 
observations \citep{Gaspari2012,Gaspari2014,Li2015}.

  \begin{table*}
 \caption{Values of $r_{500}$ and $r_{200}$ in Kpc.}
 \label{fit}
 \begin{tabular}{|l|ccccccc}
  \hline
    \cline{5-8}
&&&&\multicolumn{2}{c}{$P_{nt}=0$}&\multicolumn{2}{c}{$P_{nt}\neq0$}\\
\cline{5-8}
$cluster$&$z$   &State & $r_{500}^{planck}$& $r_{500}$   &$r_{200}$& $r_{500}$&$r_{200}$\\
  \hline 
A85  &    0.052 &CC  &  1206     &  1300 (105)  &   2319 (72)    & 1422 (147)   &2390  (51)    \\
A119 &    0.044 &NCC &  1114     &  1026 (49)   &   2070 (113)   & 1101  (64)   &2188  (46)    \\
A401 &    0.075 &NCC &  1355     &  1265 (53)   &   1883 (140)   & 1340  (61)   &2038  (182)    \\
A478 &    0.088 &CC  &  1326     &  1309 (56)   &   1923 (153)   & 1390  (65)   &2100  (214)     \\
A665 &    0.182 &NCC &  1331     &  1162 (56)   &   1846 (123)   & 1262  (69)   &2036  (144)     \\
A1651&    0.084 &NCC &  1135     &  1165 (62)   &   1946 (250)   & 1247  (81)   &2186  (61)     \\
A1689&    0.183 &NCC &  1339     &  1368 (54)   &   1877 (103)   & 1449  (60)   &2010  (125)     \\
A1795&    0.062 &CC  &  1254     &  1246 (58)   &   1864 (151)   & 1337  (69)   &2058  (202)      \\
A2029&    0.078 &CC  &  1392     &  1332 (58)   &   1989 (154)   & 1421  (68)   &2182  (207)      \\
A2204&    0.152 &CC  &  1345     &  1307 (52)   &   1877 (124)   & 1388  (60)   &2035  (161)      \\
A2218&    0.171 &NCC &  1151     &  1001 (35)   &   1496 (105)   & 1058  (43)   &1624  (139)      \\
A2255&    0.081 &NCC &  1169     &  1252 (73)   &   1827 (122)   & 1352  (82)   &1971  (143)      \\
A2256&    0.058 &NCC &  1265     &  1314 (50)   &   1781 (95)    & 1390  (56)   &1905  (116)       \\
A3112&    0.070 &CC  &  1062     &  1015 (40)   &   1459 (97)    & 1076  (45)   &1586  (132)       \\
A3158&    0.060 &NCC &  1124     &  1037 (43)   &   1521 (100)   & 1105  (48)   &1656  (133)       \\
A3266&    0.059 &NCC &  1354     &  1478 (121)  &   2592 (85)    & 1652  (166)  &2683  (35)       \\
A3558&    0.047 &NCC &  1170     &  1126 (64)   &   2017 (252)   & 1217  (83)   &2269  (49)       \\
  \hline
 \end{tabular}
 
Columns (1), (2),  (3) \& (4) shows cluster names, redshift, state, and $r_{500}$ values from \cite{Planck2011} respectively.
Columns (5) \& (6) shows values $r_{500}$ and $r_{200}$ for $P_{nt}=0$ case. Columns (7) \& (8) shows values of $r_{500}$, $r_{200}$ for $P_{nt}\neq0$ case.\newline
The numbers in brackets indicate increase in  $r_{500}$ and $r_{200}$ if clumping is taken into account.
\end{table*}

 Moreover, it has been found with  Suzaku observations that the entropy profile flattens out at large radii \citep{Hoshino2010,Simionescu2011,Eckert2013a,Fujita2013}.
This entropy decrement can be related to the gas clumping \citep{Simionescu2011,Eckert2013a, Eckert2015}, presence of non-thermal pressure \citep{Femiano2014,Su2015}, accretion/merger shocks in outskirts of clusters \citep{Hoshino2010,Cavaliere2011}, loss of kinetic energy of gas due to the cosmic ray acceleration \citep{Fujita2013, Su2015} or due to the 
rapid temperature decrease in the outskirts of clusters as a result of the non-gravitational processes  \citep{Femiano2014}.

It is convenient to define an entropy profile\footnote{Thermodynamic definition of specific entropy being $S=\ln K_g^{3/2}+$ constant.} of gas as,
$K_g(r)=kTn_e(r)^{-2/3}\propto P(r)\rho(r)^{-\gamma}$, where $k$ is the Boltzmann constant, and the exponent $\gamma=5/3$ refers to  the adiabatic index. With this definition, $K_g$ remains unchanged for all adiabatic processes and can therefore probe  the thermal history of gas. 
Purely gravitational models predict entropy profiles in clusters of the form $K_g(r) \propto r^{1.1}$ \citep{Voit2005}. 
However, as pointed, several recent observations found deviations from this expected entropy profile, especially at inner and outer radii \citep{Voit2005,10,Cavagnolo2009,Eckert2013a} as a result of non-gravitational feedback. 
To allow a meaningful interpretation and to estimate degree of feedback,  one needs to compare recent observations with theoretically expected profiles with no feedback.

Previously, \cite{Chaudhuri2012,Chaudhuri2013} estimated the non-gravitational energy deposition profile up to $r_{500}$
by comparing the observed entropy profiles with a benchmark entropy profile without feedback \citep{Voit2005} for the  REXCESS sample of 31 clusters \citep{10} observed with  XMM-Newton.  They found an excess mean energy per particle of $2.74$ keV and $1.64$ keV using benchmark entropy from adaptive mesh refinement (AMR) and smoothed particle hydrodynamics (SPH) simulations respectively, along with a strong correlation for AGN feedback. Our study extends their work
by going beyond $r_{500}$. Here, we consider the joint data set of Planck SZ pressure profile and  ROSAT gas density profiles of 17 clusters \citep{Eckert2013a,Eckert2013b,Planck2013a} to estimate the excess entropy and feedback energy profiles up to  $r_{200}$.  
Recent simulations show significant level of non-thermal pressure from bulk motion \citep{Rasia2004,Battaglia2012,Shi2015} and gas clumping (which by definition is
measured by $C$=<$\rho_g^2$>/<$\rho_g$>$^2$)  \citep{Nagai2011,Eckert2013a,Eckert2015,Battaglia2015} in the outer regions of ICM. We therefore, incorporate both these factors in our calculations.   

This paper is a continuation of our recent work \cite{Iqbal2016a} wherein we showed that pre-heating scenarios are ruled 
out at more than 3$\sigma$  statistical level. In this work, we present a detailed study of the excess entropy profiles along with feedback energy profiles and discuss the effect of non-thermal pressure and clumping in our estimates. We also look at sample selection, cool-core vs non cool-core clusters, effects of boundary conditions, choice of benchmark theoretical entropy profiles, choice of X-ray methodology, etc.

The paper is organized as follows: In section 2, we describe the cluster sample used in this work. In  section 3, we describe the self-similar non-radiative model for galaxy clusters. Section 4 is dedicated to the determination of
excess entropy and energy profiles, and the effects of the non-thermal pressure and gas clumping on their estimates. In  section 5, we check the robustness of our results. In section 6, we compare the feedback profiles for AMR and SPH benchmark entropy profiles. Section 7 gives the comparison of our results with the previous estimates. Finally, the conclusions of our work is given in last section.  Throughout this paper, we assume cosmology where ($\Omega_m$, $\Omega_{\Lambda}$, $H_0$) = (0.3, 0.7, 70).
\section{CLUSTER SAMPLE AND DATA SET}
In this work, we have studied a sample of 17 clusters in the redshift range ($0.04-0.2$) that are common in \cite{Planck2013a} and \cite{Eckert2012}. 
Based on their central entropy \citep{Cavagnolo2009}, six of these clusters are classified as cool-core ($ K_0 < 30$ keV cm$^2$), while the remaining 11 are non-cool core. 

This sample  was earlier used by \cite{Eckert2013a,Eckert2013b}\footnote{www.isdc.unige.ch/$\sim$deckert/newsite/Dominique\_Eckerts\_Hom\\epage.html.} where they have shown that the thermodynamic state of the ICM can be accurately 
recovered by using the Planck SZ pressure profile \citep{Planck2013a} and  ROSAT Position Sensitive Proportional Counter (PSPC)  gas density profile \citep{Eckert2012}\footnote{The cluster ``A2163'' from \cite{Eckert2013a,Eckert2013b} sample has been left out as its estimated feedback profile was found hugely different from others. This cluster has been found to be in a perturbed state \citep{Soucail2012}.}. 
Since the SZ signal  is proportional to the integrated pressure (gas density), unlike the X-ray signal, which is proportional to the square of density, it decreases more gently at large radius and therefore can provide more accurate pressure profiles in the cluster outskirts. This allows us to accurately recover the temperature profile beyond $r_{500}$. 
 We use parametric profiles obtained by them in this work (see \cite{Eckert2013a} for more details) which were obtained by fitting a functional form to the projected emission-measure data \citep{Vikhlinin2006} and Planck SZ pressure data \citep{Nagai2007}.  The errors were obtained through Monte Carlo Markov Chain, with direct sampling of the posterior temperature, entropy, and gas fraction distributions.

 \cite{Eckert2013a} also obtained deprojected profiles by estimating density profile through ``onion peeling'' technique \citep{Kriss1983,Eckert2012} and interpolating the SZ pressure profile. Correction for edge effects  were also applied along with median smoothing regularization for minimizing the roller-coaster effect \citep{McLaughlin1999}. The error bars were recovered by perturbing the original profile using a Monte Carlo and recomputing the deprojected profiles each time. 
 
 Since the  parametric profiles are forced to be regular, this reduces the cluster to cluster scatter and the errors. At smaller radii the angular resolution of both  Planck and ROSAT is insufficient to obtain reliable constraints. Therefore, the parametric fitting was only performed on the data beyond $0.2~r_{500}$ and were found to be consistent with the deprojected profiles.

\section{THEORETICAL MODELS}
\subsection{Cluster model}
We use the hydrostatic equation to obtain the  total mass profile $M_{tot}(r)$ in the galaxy clusters,
\begin{equation}
M_{tot}(r)\, =\,-\frac{r^2}{G\rho_g(r)}\,\frac{dP_g(r)}{dr},
\label{eqn1:eq1}
\end{equation}
where  $\rho_g$ and $P_g$ are density and thermal pressure of the ICM respectively. $dP_g(r)/dr$ is calculated by using the best fit generalized NFW (GNFW) pressure profile \citep{Planck2013a}.
The quantities $r_{500}$ and $r_{200}$ where obtained by first interpolating the $M_{tot}(r)$ profile and then iteratively solving\footnote{$\Delta$ is defined such that $r_{\Delta}$ is the radius out to which mean matter density is $\Delta \rho_c$, where $\rho_c=3H^2(z)/8\pi G$ 
being critical density of the universe at redshift $z$.},
\begin{equation}
m_{\Delta}\,=\,4/3\,r^3_{\Delta}\,\Delta \,\rho_c(z).
\label{m500}
\end{equation}
The virial radius, $r_{vir}(m_{vir},z)$, is calculated with the help of the spherical collapse model \citep{Bryan1998}.
If required, the mass profile is obtained by linear  extrapolation in  logarithmic space.
\begin{figure}
\begin{minipage}{9.7cm}
 \includegraphics[width = 9.7 cm]{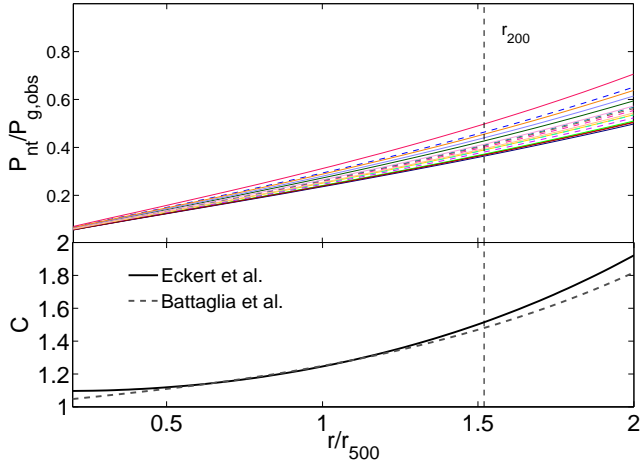}
\end{minipage}
\caption{Upper panel: Variation of $P_{nt}/P_{g}$ as a function  $r/r_{500}$ for all the clusters. $P_g$ is obtained by using the best fit GNFW pressure profile from \citet{Planck2013a}. 
 Solid lines represent NCC clusters and dashed lines represent CC clusters. Lower panel: Comparison between \citet{Eckert2015} and \citet{Battaglia2015} clumping profiles for the average case.
}
\label{pressureanddensity}
\end{figure}

Recent simulations suggest that a significant non-thermal pressure contributes to the total energy of the ICM gas, mainly due to bulk gas motions 
and turbulence in the ICM gas \citep{Vazza2009,Battaglia2012,Shi2015}. 
While non-thermal pressure is small in the inner region, its relative importance steadily increases with radius, becoming a significant fraction of the total pressure in the outer region \citep{Lau2009}. It has been found from both observations and simulations that $m_{500}$ is underestimated by about $10-20\%$, if one uses the hydrostatic equation without non-thermal pressure \citep{Rasia2004,Shi2015}.
From the recent numerical simulations, we model the non-thermal pressure fraction in the power law form similar to that given in \cite{Shaw2010},
\begin{equation}
 P_{nt}(r,z)\,=\,f(r,z)\,P_{tot}\,=\,\frac{f(r,z)}{1-f(r,z)}\,P_g(r),
 \label{eqn2:eq2}
\end{equation}
where $P_{tot}$ is the total gas pressure, $f(r,z)=a(z)\left(\frac{r}{r_{500}}\right)^{n_{ nt}}$, $a(z)=a_0(1+z)^{\beta}$ for low redshift clusters ($z\le1$) with $a_0=0.18\pm0.06$, $\beta=0.5$ and $n_{nt}=0.8\pm0.25$ \citep{Shaw2010}.
In upper panel of Fig.~\ref{pressureanddensity}, we have plotted $P_{nt}/P_{g}$ as a function  $r/r_{500}$. It can be seen that $P_{nt}$ becomes  $\sim50\%$ of the thermal gas pressure $P_{g}$ in the cluster outskirts. 
Since non-thermal pressure is not negligible beyond $r_{500}$, one should take it into account in order to properly study the cluster physics in the outer regions. 

Similarly, it has been seen that gas clumping results in an overestimation of the observed gas density ($\rho_{g,obs}$) and hence underestimation of the entropy and total mass profiles. It has been found from various observations and hydrodynamical simulations that the clumping factor is negligible in the innermost cluster regions but radially increases with $\sqrt C\approx1-2$ around $r_{200}$ \citep{Mathiesen1999,Nagai2011,Vazza2013,Battaglia2015}. However, few works have also reported either smaller or higher values of clumping factor \citep{Walker2013,Femiano2013,Urban2014,Femiano2014}. 
\cite{Eckert2015} found that the azimuthal median is a good tracer of the true 3D density ($\rho_{g,true}$) and showed from both  hydrodynamical simulations that their method recovered the $\rho_{g,true}$ profiles with  deviations less than 10\% at all radii. They recovered the average  $\sqrt C=1.25$ at $r_{200}$ which is consistent with the recent results. Lower panel of Fig.~\ref{pressureanddensity}  shows the consistency of \cite{Eckert2015} clumping profile with that of \cite{Battaglia2015}.

We have also calculated the total mass profile $M_{tot}$ by including non-thermal pressure $P_{nt}$ in Eq.~\ref{eqn1:eq1} and correcting the density profile using  \cite{Eckert2015} clumping profile.
In Tab.~\ref{fit}, we give estimates  of $r_{500}$ and $r_{200}$ obtained by using parametric  density profiles along with {\it Planck} best fit GNFW pressure profile \citep{Planck2013a}. 
For comparison, we have also given Planck $r_{500}$ values from \cite{Planck2011} which are consistent with our estimates within $10\%$  for most of the clusters.
Moreover, we find that the average scaling $r_{200}=1.52~r_{500}$ \citep{Pointecouteau2005,11,Eckert2013a} is in excellent agreement with our results with small scatter.
\begin{figure}
\begin{minipage}{8.5cm}
\includegraphics[width=8.5cm]{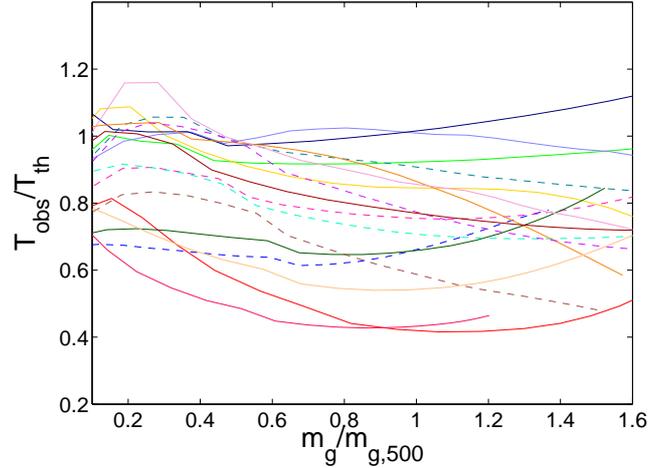}
\end{minipage}
\caption{This plot shows variation of $T_{obs}/T_{th}$ as a function  $m_g/m_{g,500}$ for all the clusters (considering clumping and $P_{nt}\neq 0$). Solid lines represent NCC clusters and dashed lines represent CC clusters.}
\label{Tobs_by_Tth}
\end{figure} 

\begin{figure*}
\begin{minipage}{8.5cm}
 \includegraphics[width = 8.5 cm]{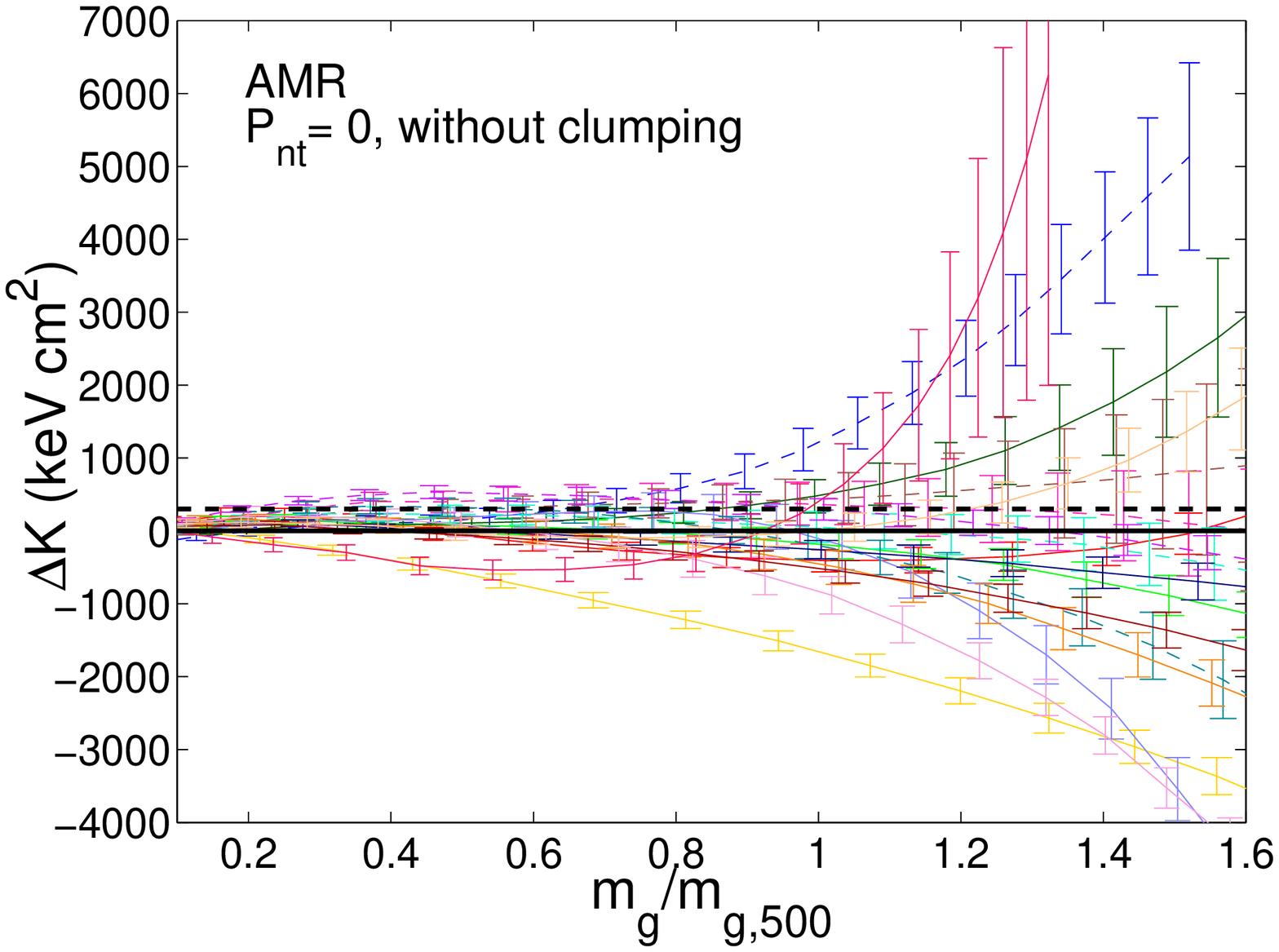}
\end{minipage}
\begin{minipage}{8.5cm}
 \includegraphics[width = 8.5 cm]{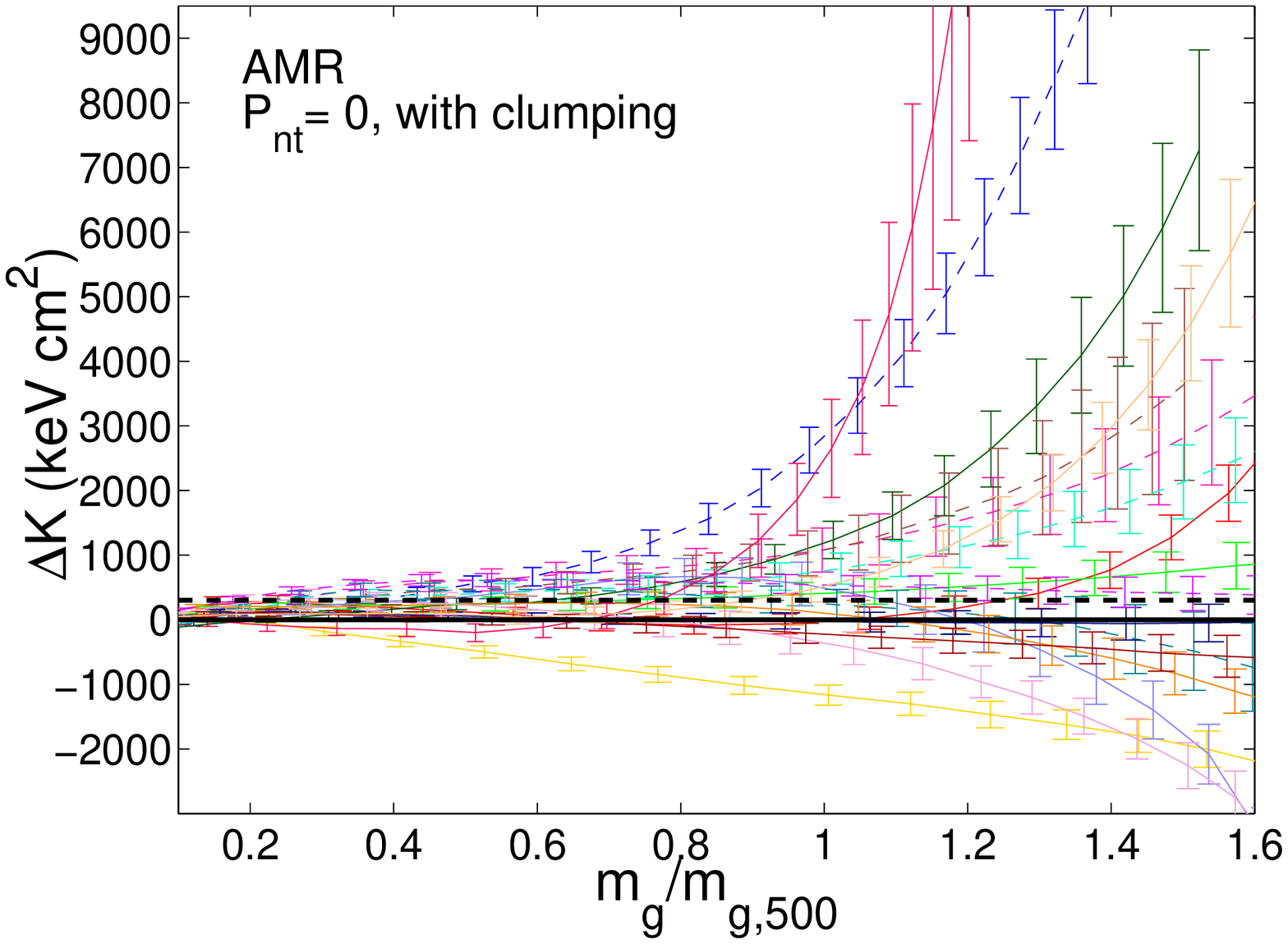}
\end{minipage}  
\begin{minipage}{8.5cm}
 \includegraphics[width = 8.5 cm]{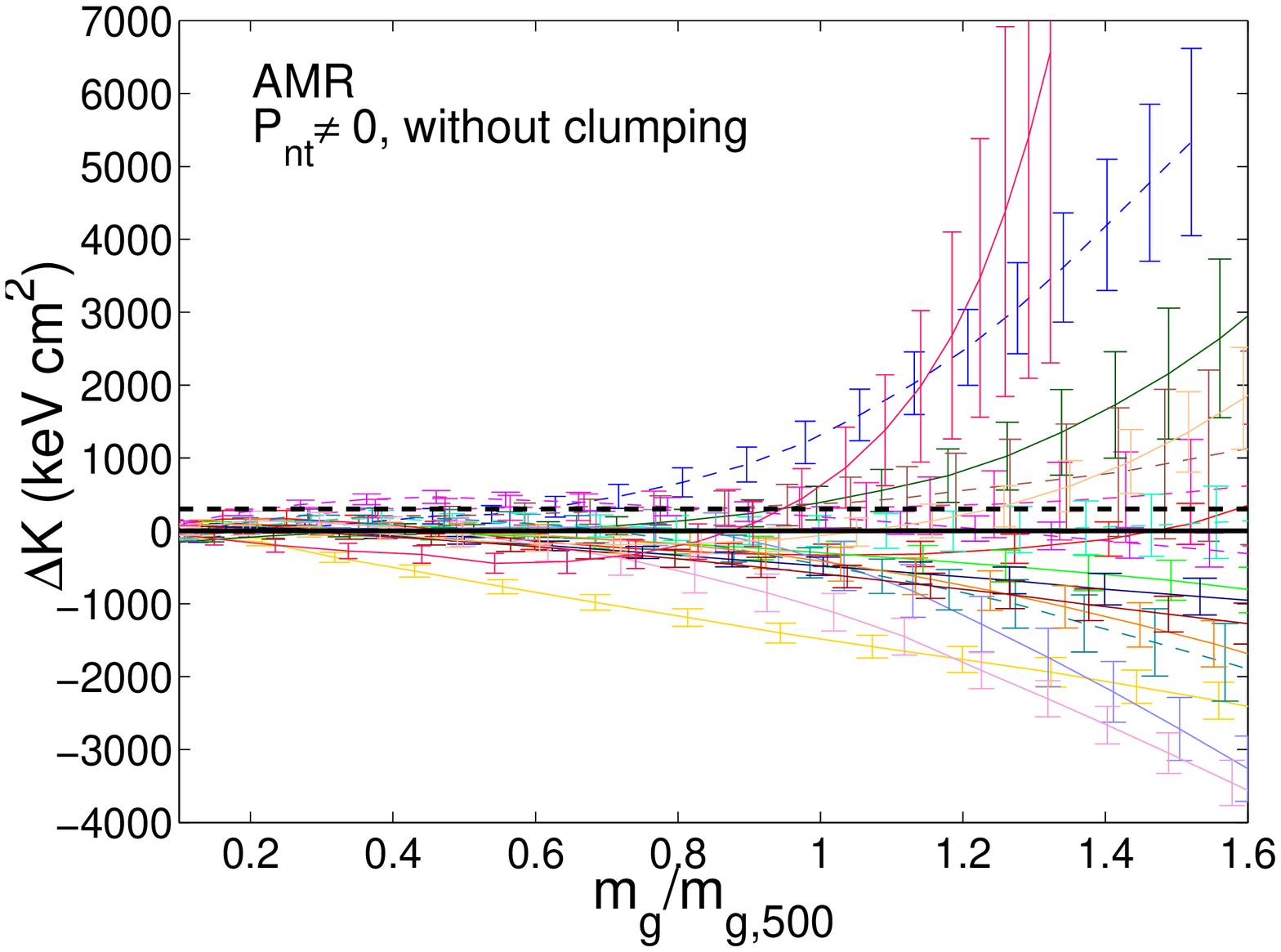}
\end{minipage}
\begin{minipage}{8.5cm}
 \includegraphics[width = 8.5 cm]{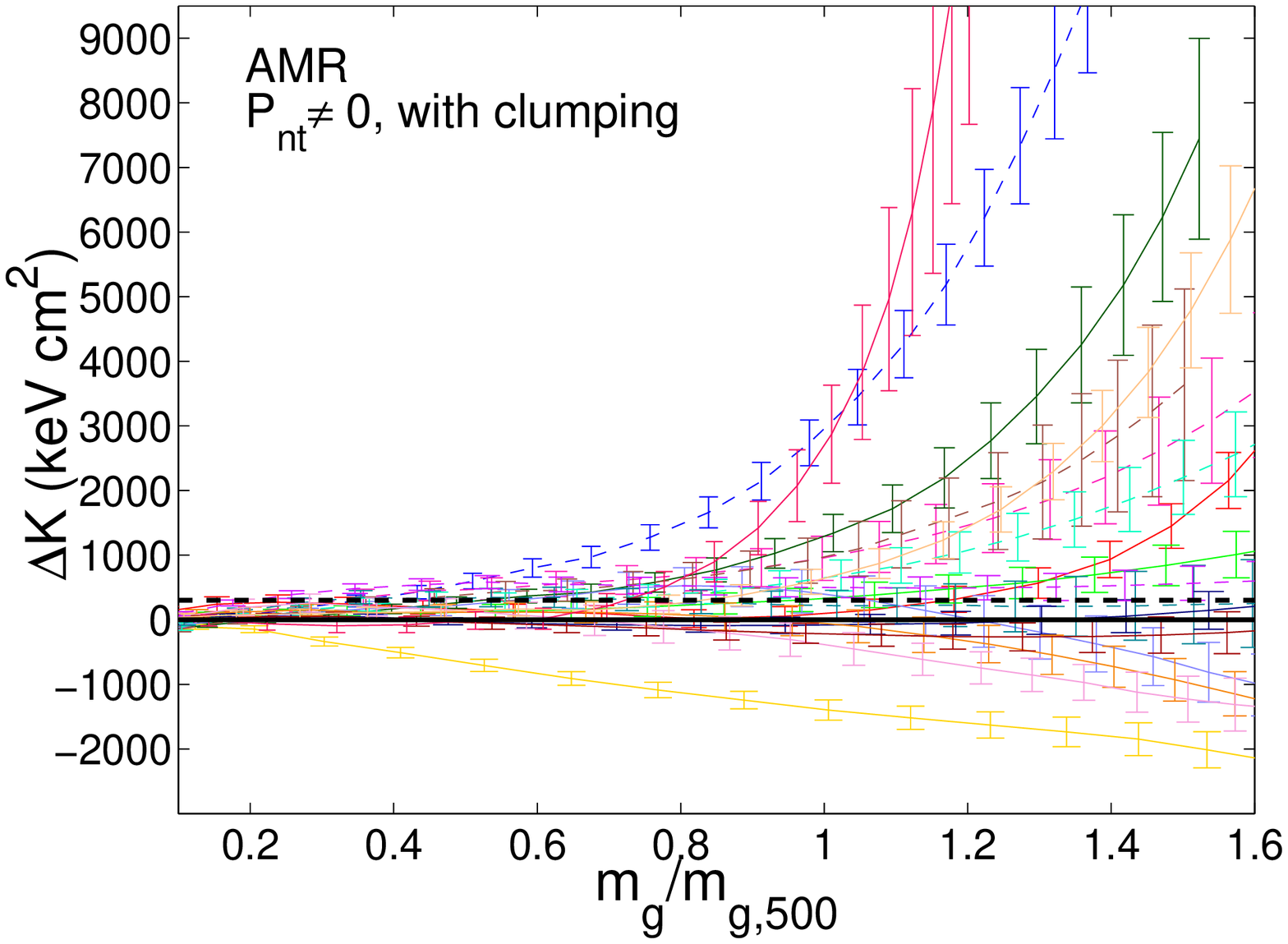}
\end{minipage}   
\caption{
Excess entropy $\Delta K$  as a function  $m_g/m_{g,500}$ using benchmark AMR entropy profile.
Left panel: without clumping, Right panel: with clumping. Upper panel: $P_{nt}=0$, Lower panel: $P_{nt}\neq0$. Thin dashed lines represent NCC clusters and dashed red lines represent CC clusters. The error bars are given at  1$\sigma$ level.
Note that for meaningful comparison, we have scaled x-axis of all cases with same $m_{g,500}$ as that of fiducial case (i.e with clumping  and $P_{nt}\neq0$).}
\label{fig:AmrdeltaK}
\end{figure*}
 
\subsection{Initial entropy profile}
Standard models of large scale structure show that matter is shock heated as it falls into clusters under the influence of gravity and predict that the entropy of gas ($K_{g,th}$) should behave as a
power-law, with entropy profiles flattening near cluster cores. \cite{Voit2005} performed several non-radiative SPH and AMR simulations in order to study the main features of the 
entropy profiles. They found differences in entropy profiles in the inner cores but the differences were small for $ r>0.2~r_{200}$ in SPH and AMR simulations. They found that the simulated non-radiative scaled entropy profile can  be  described by a simple power law form in the range    $(0.2-1)~r_{200}$ with  a slightly higher normalization for AMR case.
For the inner radii, \cite{Voit2005}  found a large discrepancy in the scaled entropy profiles between the SPH and AMR simulations. A flat entropy core has been observed in the center of non-radiative galaxy clusters in Eulerian grid codes (AMR) which is absent in Lagrangian approaches (SPH). However, after accounting for certain hydrodynamical processes (i.e shocks and mixing motions)  
the results of SPH simulations match with that of AMR case \citep{Mitchell2009,Vazza2011,Power2014}.
We use AMR and SPH median entropy profiles obtained by \cite{Voit2005} for our baseline model described by,
\begin{equation}
\frac{K_{g,th}}{K_{200}}=a_0\left(\frac{r}{r_{200}}\right)^{1.1},
\label{eq:voit1}
\end{equation}
in the range $(0.2-1)~r_{200}$ plus by a flatter core below $0.2~r_{200}$ which is much more pronounced in case of AMR simulations.  $a_0$ is equal to 1.32 and 1.41 for SPH and AMR respectively and $K_{200}$ is given by,
\begin{equation} 
K_{200}=144 \left(\frac{m_{200}}{10^{14}M_\odot}\right)^{2/3} \left(\frac{1}{f_b}\right)^{2/3} h(z)^{-2/3} \mbox{ keV cm}^2.
\label{K200}
\end{equation}
where $f_b$ is the universal baryonic fraction and $h(z)=H(z)/H_0$, $H(z)$ being hubble constant at redshift $z$.

In order to calculate the initial (without feedback) density (or gas mass) and temperature profiles, one solves the the hydrostatic equation with appropriate boundary condition \citep{Chaudhuri2012,Chaudhuri2013}. 
Considering non-thermal pressure component, we rewrite the hydrostatic equation as,
\begin{equation}
\frac{d(P_{g,th}+P_{nt,th})}{dr} =-\left( \frac{P_{g,th}}{K_{g,th}} \right )^{3/5}  m_p \mu_e^{2/5} \mu^{3/5}  \frac{ G M_{tot}( <r ) }{ r^2} ,
\label{H:he2}
\end{equation}
where $P_{g,th} = n_{g,th}kT_{th}$ is the initial (theoretical) thermal pressure of the ICM, $T_{th}$ is the initial ICM temperature and $M_{tot}$ is the sum of two terms, $M_{tot}=M_{thermal}+M_{non-thermal}$. 
Since energy injection only effects the gas mass profile, one can assume the dark matter profile and hence total mass profile to remain constant during the feedback processes.
For the boundary condition  we assume the gas fraction ($f_{g,th}$) to be $0.9f_b$ at virial radius \citep{Crain2007}.
On the addition of non-thermal pressure, the initial entropy profile is increased due the overall increase in the normalization and therefore, the  deviation from the observed entropy decreases. 

It is important to note that initial entropy profile also depends on the baryonic fraction through $K_{200}$. Most of the previous estimates of the entropy floor were based on the 
WMAP7 estimates of $f_b=0.167$ and since the Planck predicts relatively lower value of $f_b=0.156$ \citep{Planck2013b,Planck2015}, this will further increase the initial entropy profile thereby decreasing the estimates of excess entropy. 
\subsection{Estimates of total feedback energy}
In  this section, we estimate the total mechanical feedback energy. It is important to note that for  a meaningful interpretation, one should  compare the theoretical and observed entropy profiles at the same gas mass  ($m_g$) instead of 
same radii in order to provide an allowance for redistribution of gas on account of feedback processes \citep{Li2010,Nath2011,Chaudhuri2012,Chaudhuri2013}. 
Considering a transformation from the baseline configuration to new configuration 
i.e $\Delta K(m_g)=K_{g,obs}(m_g)-K_{g,th}(m_g)$, the additional thermal energy per particle in ICM corresponding to the transformation is given by,
\begin{eqnarray}
\Delta Q_{ICM} &=& {kT_{obs} \over
(\gamma-1 )} {\Delta K \over K_{g,obs}}  \qquad\qquad \quad ({\rm isochoric})\nonumber\\
&=&{kT_{obs}  \over (1-{1 \over \gamma})} 
{  \beta ^{2/3} (\beta -1) \over (\beta^{5/3}-1)}
{\Delta K \over K_{g,obs}}  ({\rm isobaric}),
\label{eq:delq}
\end{eqnarray}
where $\beta=T_{obs}/T_{th}$. For a value of $\beta=2$, the ratio between $\Delta Q_{ICM}$ in two cases is $1.14$. This implies that if the
observed temperature $T_{obs}(m_g)$ deviates from the theoretically calculated value $T_{th} (m_g)$
by a factor $\le 2$, then the two above mentioned estimates of energy input per unit mass differ by only a factor of $1.2$. 
Fig.~\ref{Tobs_by_Tth} shows the ratio  $\beta=T_{obs}/T_{th}$ for all the clusters and is mostly in the range $0.5<\beta<1.2$. We choose the expression for 
the isobaric process in our estimates. Moreover, we find using isochoric expression instead does not make any notable difference.
The total excess energy per particle in ICM can be found by including the change in potential energy term in last equation,
\begin{equation}
\Delta E_{\rm ICM}=\Delta Q_{ICM}+G\mu m_p\left(\frac {M_{tot}(r_{th})}{r_{th}}-\frac{M_{tot}(r_{obs})}{r_{obs}}\right), 
\end{equation}
where $r_{th}$ and $r_{obs}$ are theoretical and observed radii respectively enclosing the same gas mass.

The total feedback energy per particle in ICM can be found after adding the energy lost due to cooling i.e,
\begin{equation}
\Delta E_{feedback}= \Delta E_{ICM}+ \Delta E_{cool}.
\label{energy}
\end{equation}
We approximate the energy lost in ICM in a given mass shell as,
\begin{equation}
 \Delta E_{cool}=\Delta L_{bol}\,t_{age},
\end{equation}
where $\Delta L_{bol}$ is the bolometric luminosity emitted by the ICM in a given shell which is obtained by considering cooling function $\Lambda_{N}$ given in \cite{Tozzi2001} and $t_{age}$ is the age of 
the cluster which we have fixed at $5$ Gyr \citep{Chaudhuri2013}. $\Lambda_{N}$ is calculated using theoretical (initial) temperature and density profiles. We found using observed profiles instead of theoretical profiles does not make any notable difference in our estimates. 

The total amount of energy deposited, for the whole cluster is,  
\begin{equation}
E_{feedback} \,= \,\int \Delta E_{feedback}\frac{1}{\mu_g m_p}  \, dm_g \,,
\end{equation}
where $\mu_g = 0.6$ is the mean molecular weight of gas and $m_p$ is mass of proton. Dividing the total energy in the ICM  by the total number of particles in the ICM, we estimate the average energy  per particle ($\epsilon_{feedback}$).

\begin{figure*}
\begin{minipage}{8.5cm}
 \includegraphics[width = 8.5 cm]{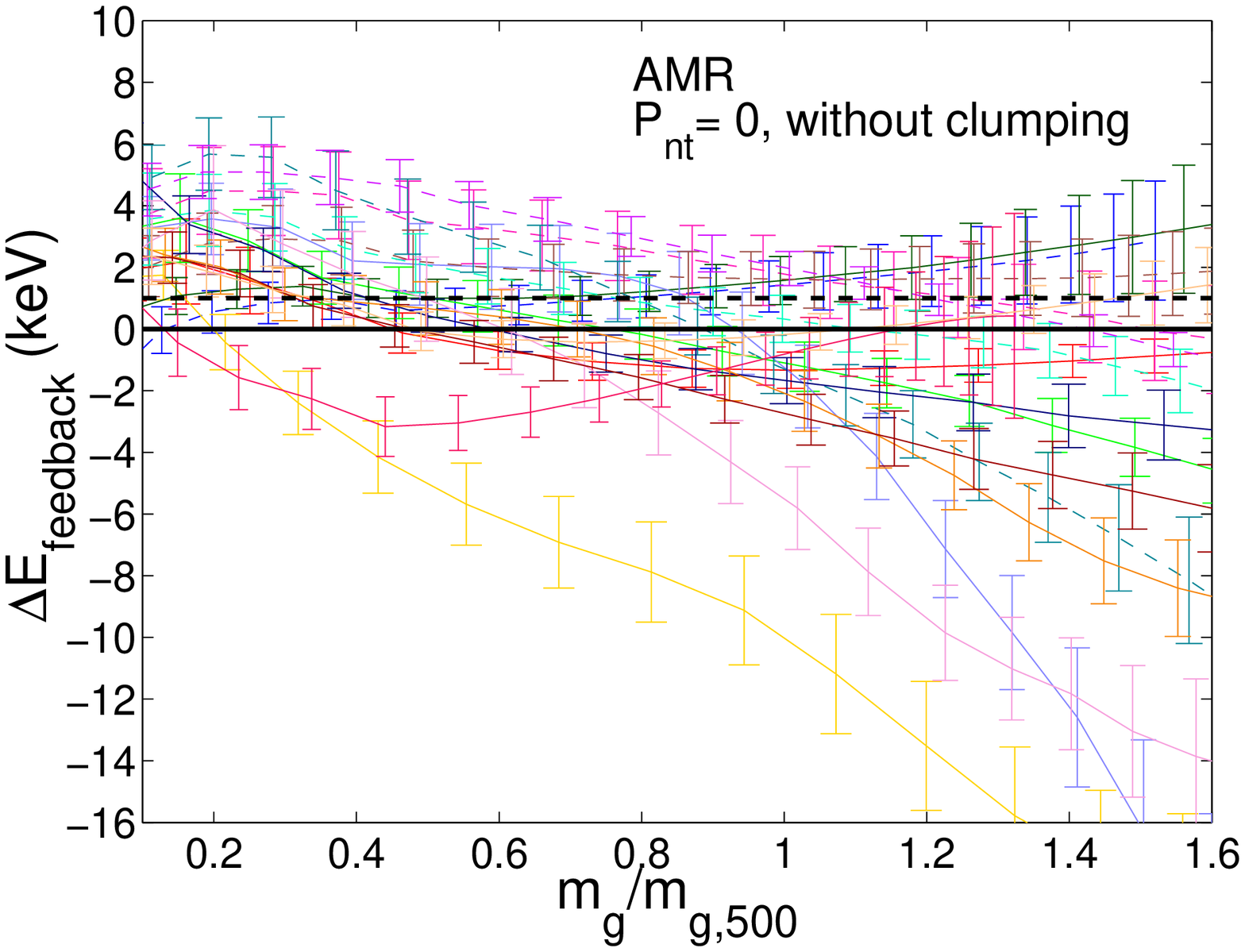}
\end{minipage}
\begin{minipage}{8.5cm}
 \includegraphics[width = 8.5 cm]{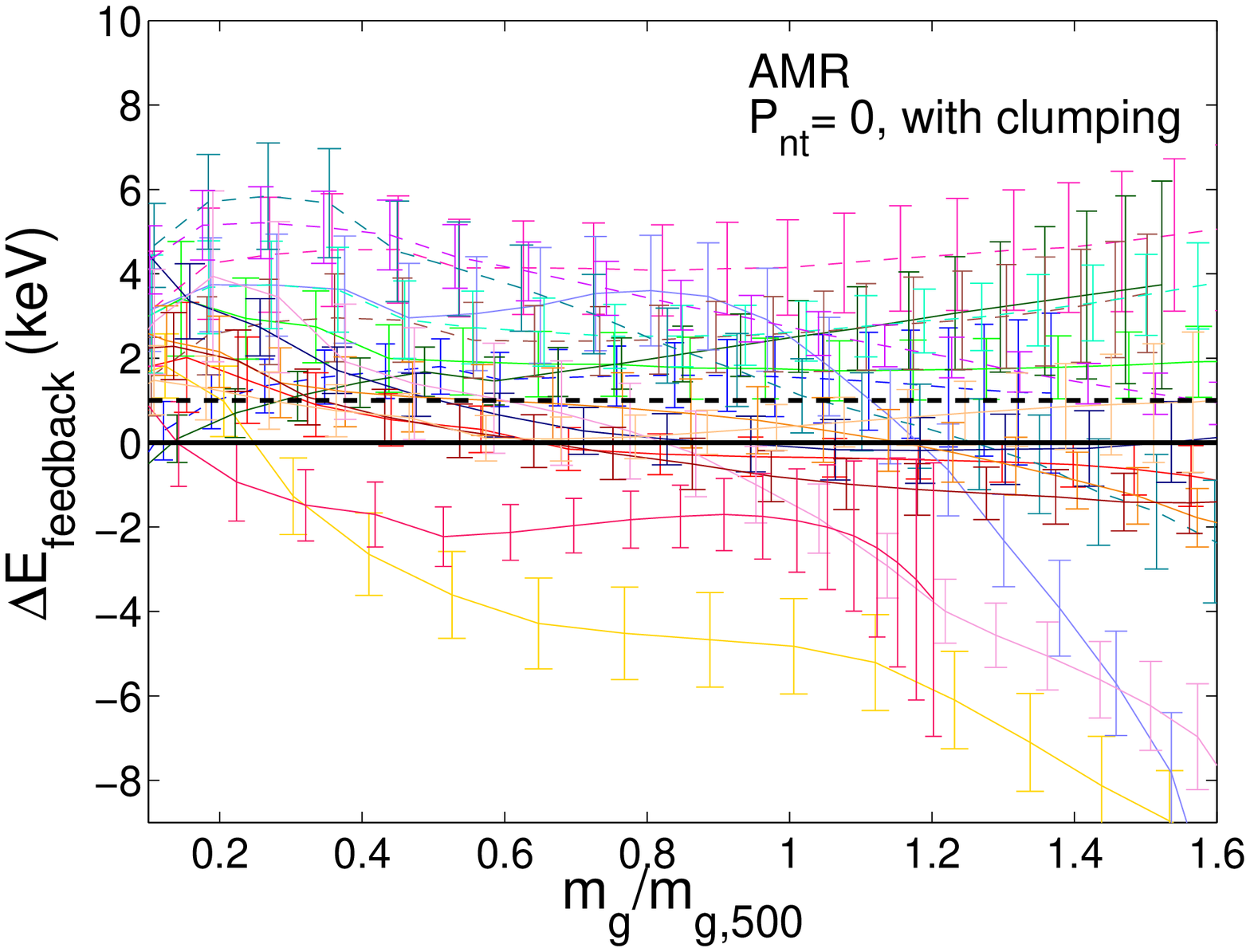}
\end{minipage}  
\begin{minipage}{8.5cm}
 \includegraphics[width = 8.5 cm]{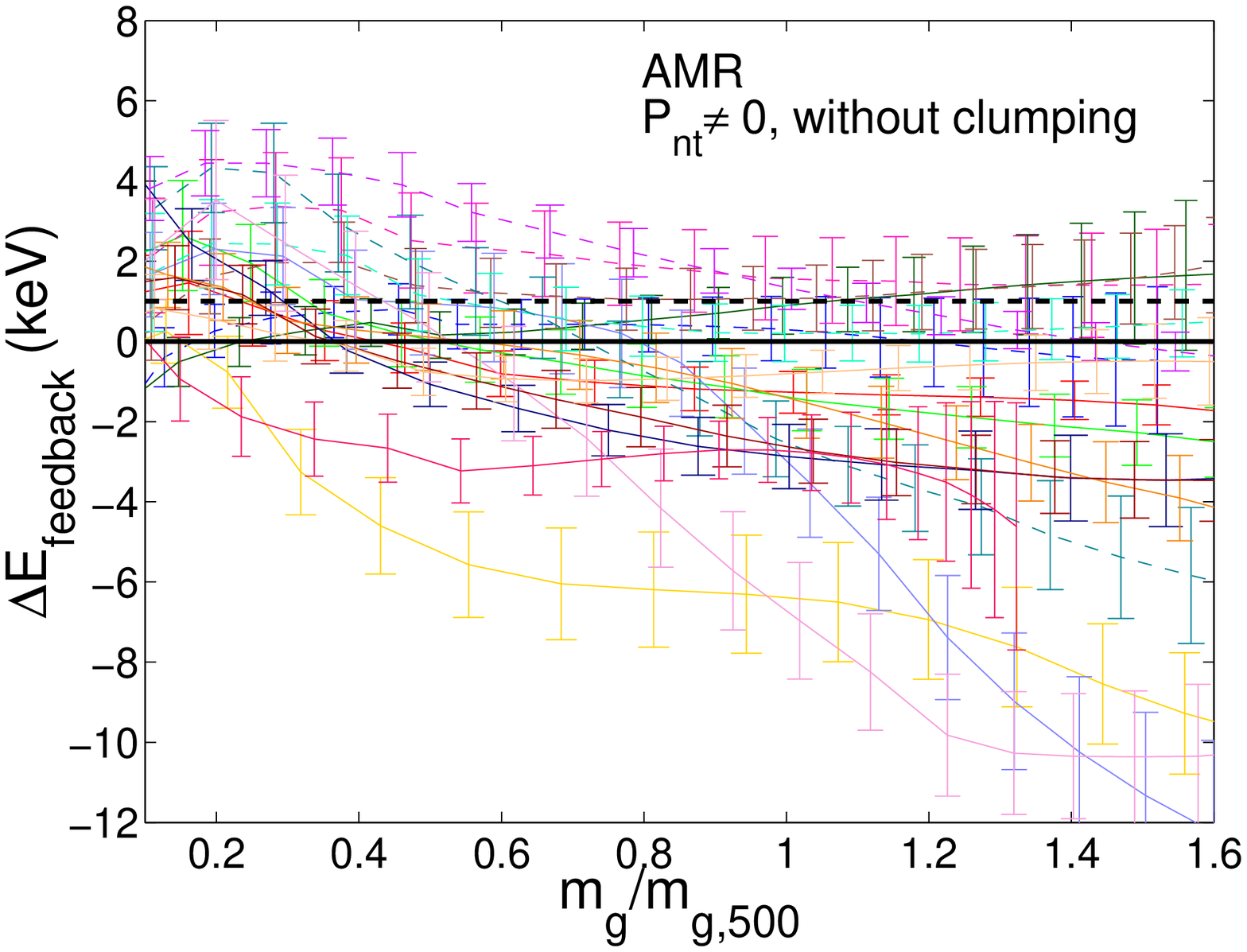}
\end{minipage}
\begin{minipage}{8.8cm}
 \includegraphics[width = 8.8 cm]{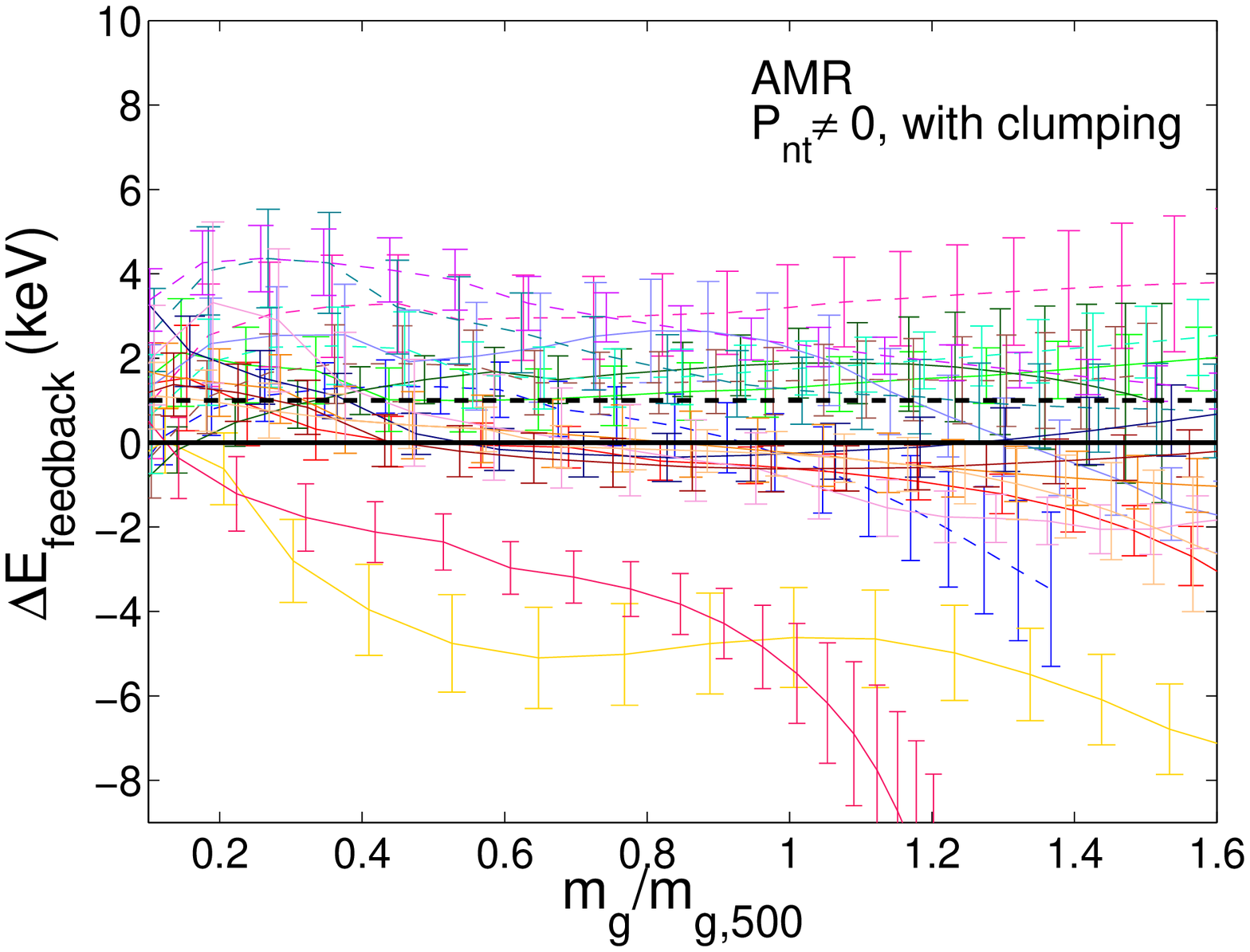}
\end{minipage}   
\caption{
Excess energy per particle $\Delta E$  as a function  $m_g/m_{g,500}$ using benchmark AMR entropy profile.
Left panel: without clumping, Right panel: with clumping. Upper panel: $P_{nt}=0$, Lower panel: $P_{nt}\neq0$. Thin solid lines represent NCC clusters and dashed lines represent CC clusters. The error bars are given at  1$\sigma$ level.
Note that for meaningful comparison, we have scaled x-axis of all cases with same $m_{g,500}$ as that of fiducial case (i.e with clumping  and $P_{nt}\neq0$).
}
\label{fig:AmrdeltaE}
\end{figure*}

\section{Results and Discussion}
In this section, we study the entropy and energy deposition profiles (in terms of $\Delta K$ and $\Delta E_{feedback}$ profiles) in the galaxy clusters up to  $r_{200}$ ($m_g/m_{g,500}=1.6$) using the methodology discussed in the previous section. We also investigate the impact of the non-thermal pressure, gas clumping and baryonic fraction on our estimates. All the figures that follow assume AMR 
entropy profiles and Planck estimates of the universal baryonic fraction $f_b=0.156$ unless stated otherwise. We shall refer to the case where we assume $f_b=0.156$ and consider both non-thermal pressure and clumping as a fiducial case. The results obtained using SPH baseline entropy profiles are shown in appendix \ref{sec:appendix}. 

\subsection{$\Delta K$ and $\Delta E$ profiles}
In Fig.~\ref{fig:AmrdeltaK}, we show $\Delta K$ profiles of all the individual clusters along with weighted average of full sample as a function of $m_g/m_{g,500}$ and Fig.~\ref{fig:AmrdeltaE} shows the corresponding $\Delta E_{feedback}$  profiles. 

In general, we find that for both $\Delta K$ and $\Delta E_{feedback}$,
\begin{itemize}
\item The average profiles for full sample are positive in the inner regions, but it becomes negative in the outer regions.
\item The inclusion of clumping factor, increases overall profiles, due to the increase of observed entropy profiles. 
\item The inclusion of non-thermal pressure decreases overall profile up to $r_{500}$ and in the outer radii the profiles actually unexpectedly increase.
\item The average profiles of CC and NCC clusters differ significantly. CC clusters have much higher values compared to the average.
\end{itemize}


\begin{figure}
\begin{minipage}{8.5cm}
\includegraphics[width=8.5cm]{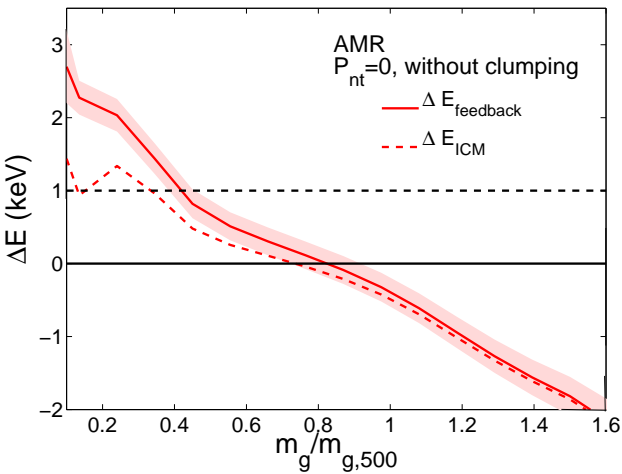}
\end{minipage}
\caption{The average $\Delta E$ profile for the entire sample with and without adding energy lost due to cooling.}
\label{Cooling}
\end{figure} 


\begin{figure*}
\begin{minipage}{8.5cm}
 \includegraphics[width = 8.5cm]{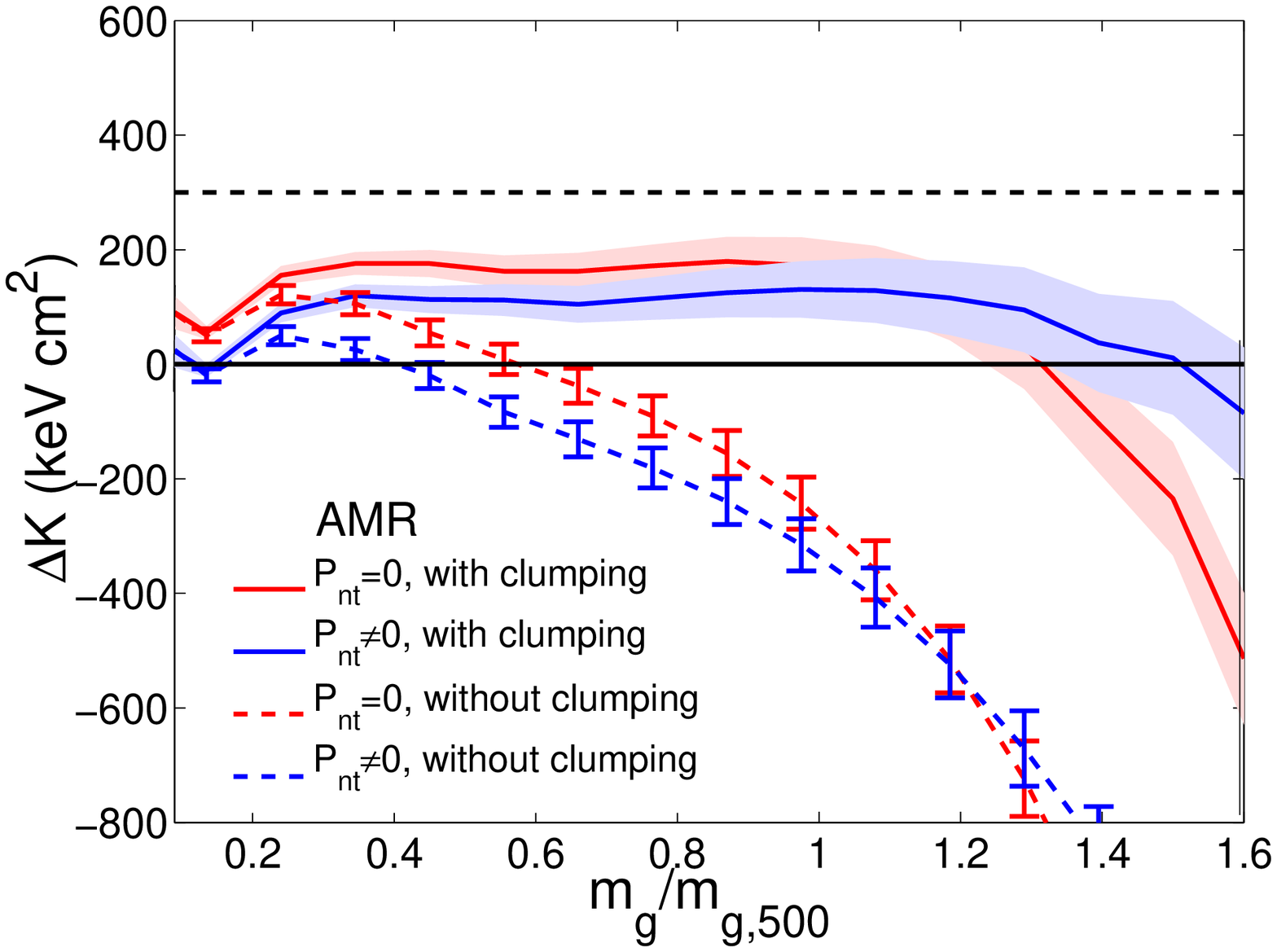}
\end{minipage}
\begin{minipage}{8.5cm}
 \includegraphics[width = 8.5cm]{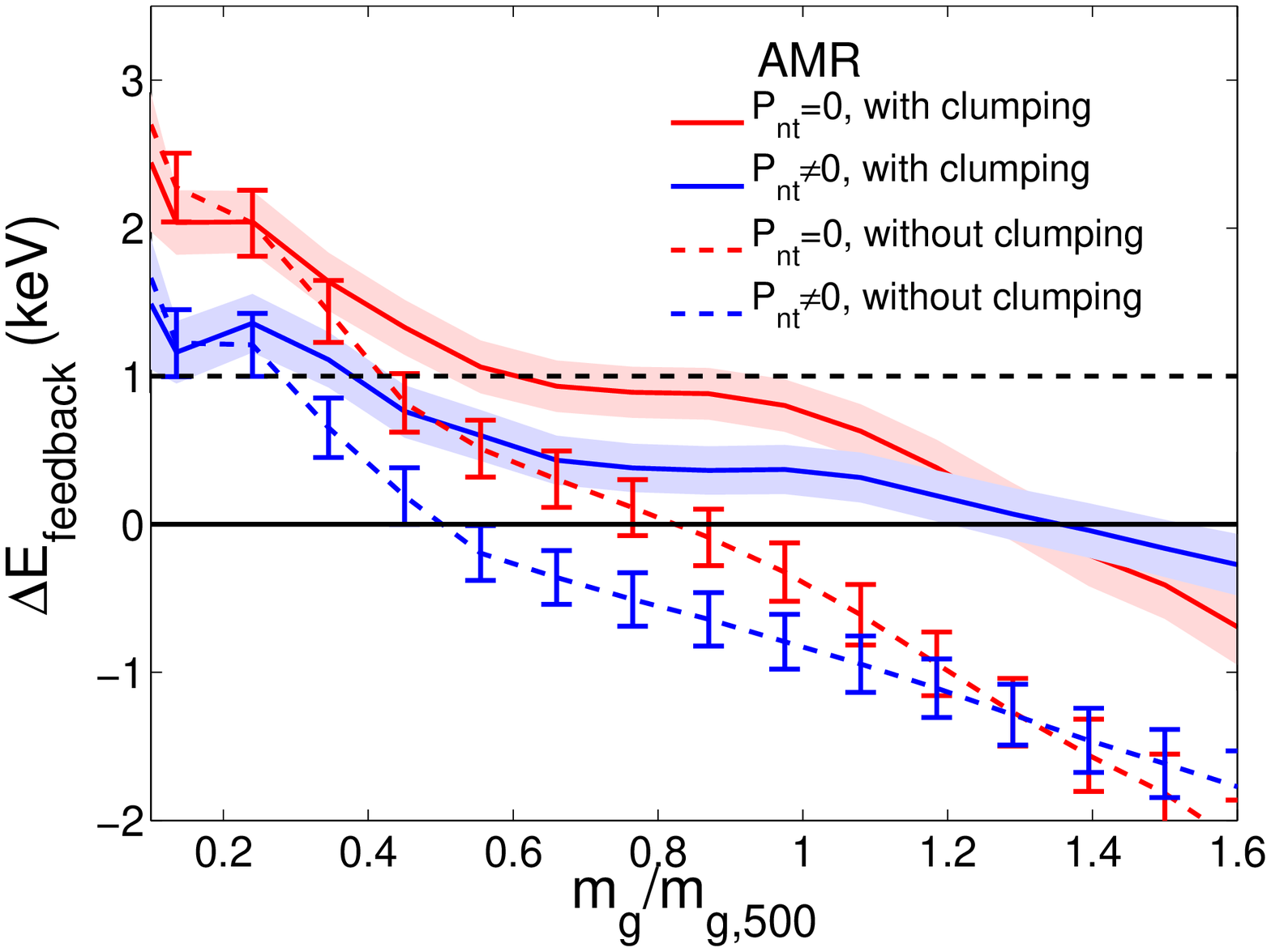}
\end{minipage}   
\caption{Comparison of average $\Delta K$  and $\Delta E$ profiles for different cases considered in this work. 
The error bars are given at 1$\sigma$ level.}
\label{fig:average}
\end{figure*}

Given the negative values of $\Delta K$ at the outer radii,  the corresponding profiles of $\Delta E$ per particle would also be negative. 
On the face of it, the result would be physically meaningless. However, one should note that ICM gas loses energy due to radiation and the amount of energy lost due to radiation can be added to offset offset the negative values. Solid red line and dashed red line in Fig.~ \ref{Cooling} 
shows the average profiles with and without taking into account energy lost due to cooling respectively. The difference in these two curves  is small beyond $r_{500}$ because of the fact that gas density is small in those regions and therefore radiative cooling cannot explain the profiles going below zero in the  outer region.  

Tabs.~\ref{amrplanck}~\&~\ref{amrwmap} give the estimates of average feedback energy per particle $\epsilon_{feedback}$ in the ranges $0.2~r_{500}-r_{500}$, $0.2~r_{500}-r_{200}$ and $r_{500}-r_{200}$ using  Planck and  WMAP estimates of the baryonic fraction. 
It can be seen that the inclusion of non-thermal pressure affects the estimates of $\epsilon_{feedback}$ both in the inner and outer regions of the cluster and that clumping has a substantial effect only in the cluster outer regions. In the next two subsections, we show as to how the proper incorporation of both clumping and non-thermal pressure can lead to meaningful estimates of the feedback profiles.
\begin{figure*}
\begin{minipage}{8.5cm}
 \includegraphics[width = 8.5cm]{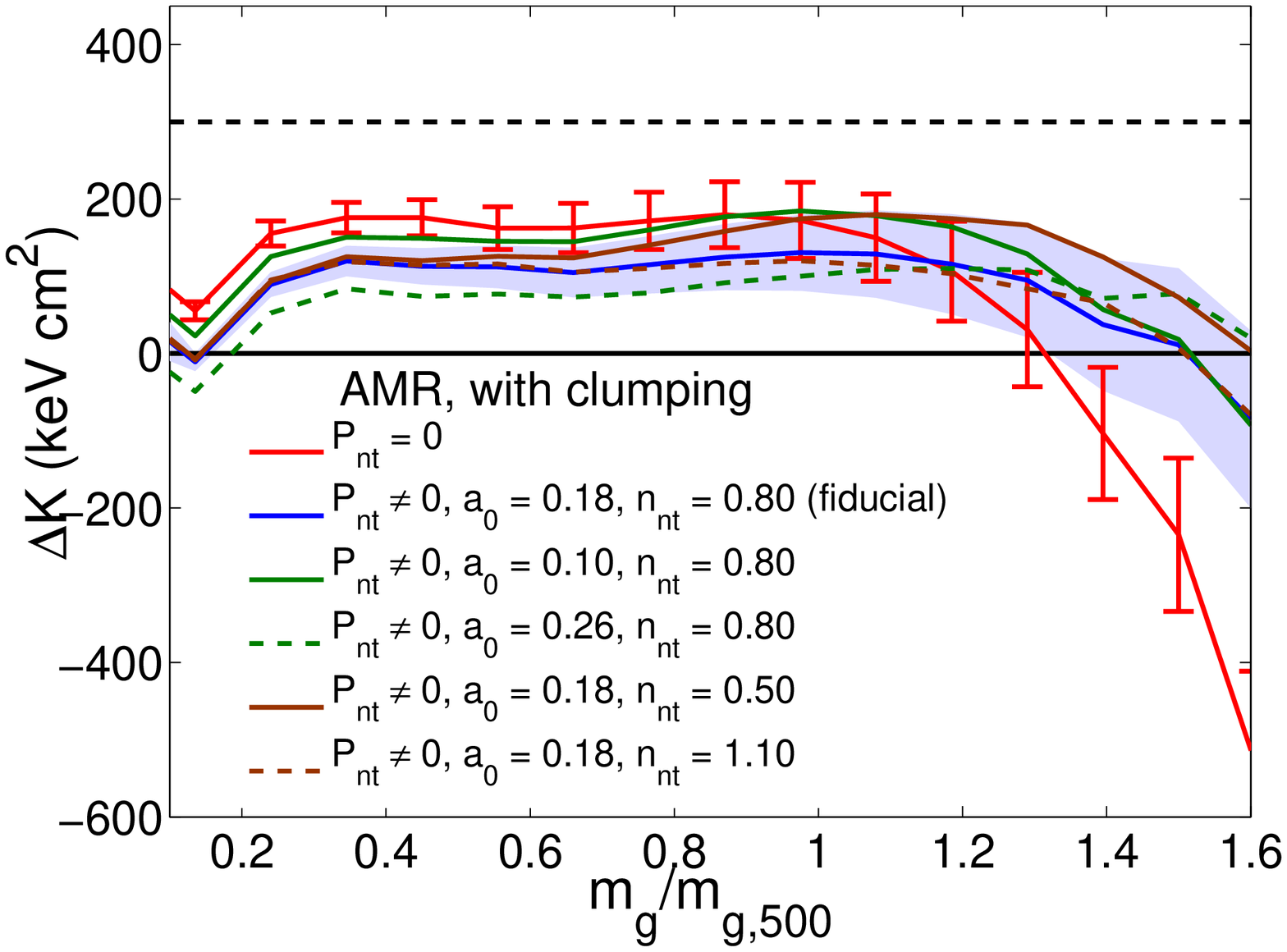}
\end{minipage}
\begin{minipage}{8.5cm}
 \includegraphics[width = 8.5cm]{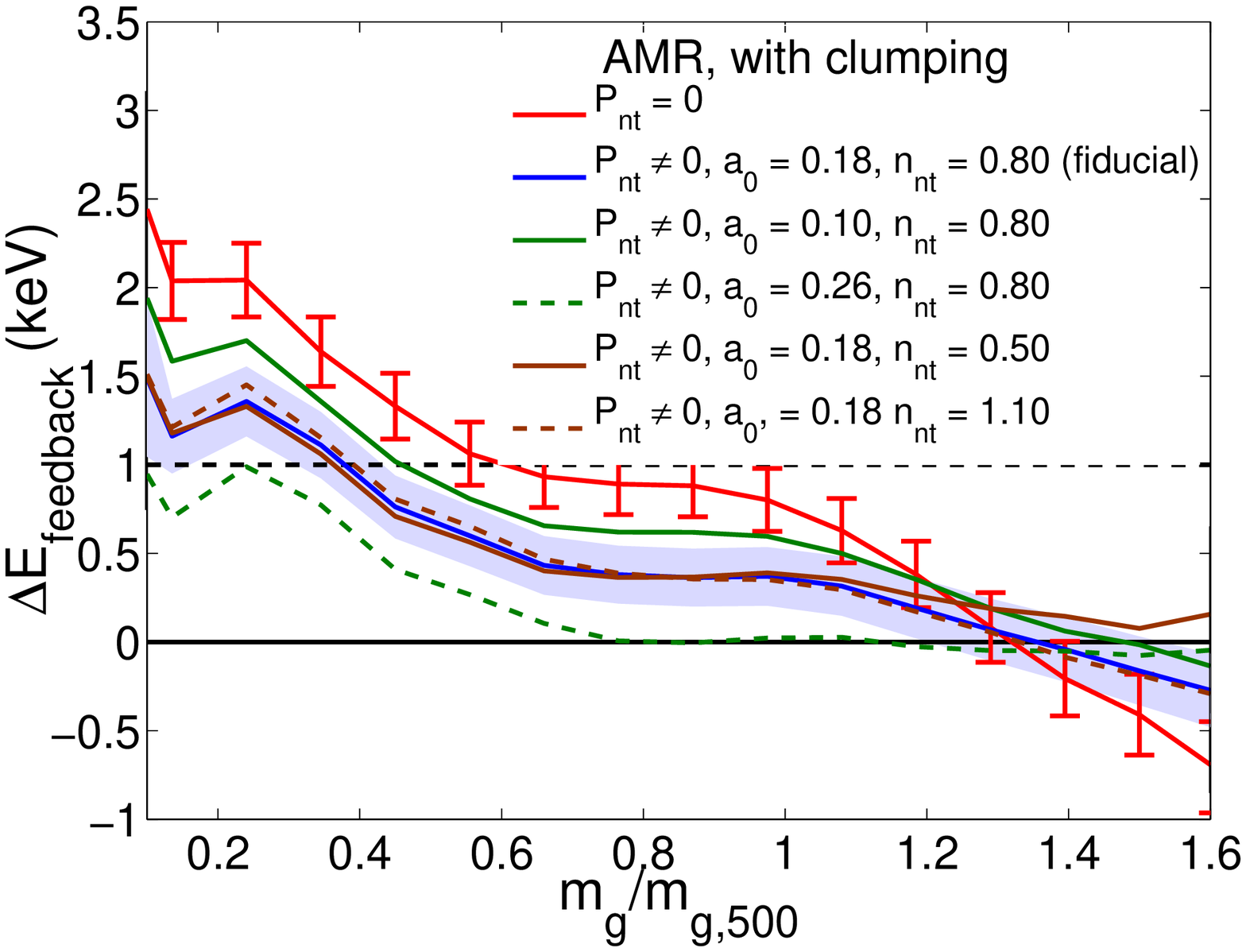}
\end{minipage}   
\caption{Comparison of feedback profiles for different parameterizations of $P_{nt}$. The error bars are given at  1$\sigma$ level.}
\label{nonthermalnor}
\end{figure*}
\subsection{Importance of gas clumping}
In the outer regions the level of clumping in gas profile can be significant which can also lead to  biased estimates
of density and hence in entropy measurements. Fig.~\ref{fig:average} shows the comparison of average $\Delta K$ and $\Delta E_{feedback}$ profiles for various cases. One can clearly see the implications of correcting the entropy by using the  clumping profile from \cite{Eckert2015} on our estimates. The addition of clumping factor raises $\Delta K$ and hence $\Delta E_{feedback}$ profiles as expected. Here, the increase is not negligible and makes the average profile become more or less consistent with zero in the outer regions for $P_{nt}\neq0$ case. However, for the pure thermal case, the profiles are still negative in the outer regions to a significant level which we ascribe to the  neglect of non-thermal pressure in the next subsection.

Comparing the average $\Delta K$ feedback profiles with and without clumping, one can see that ignoring the clumping correction leads to a decrease of $\Delta K\approx 300$ keV cm$^2$ at $r_{500}$ and $\Delta K\approx 1100$ keV cm$^2$ at $r_{200}$. The energy feedback profiles on other hand are under-estimated by $\Delta E_{feedback} \approx 1$ keV at $r_{500}$ and $\Delta E_{feedback} \approx 1.5$ keV at $r_{200}$. Similarly, from Tabs.~\ref{amrplanck}~\&~\ref{amrwmap}, it is also evident that the the average feedback energy per particle, $\epsilon_{feedback}$, is under-estimated by $0.5$ keV in the region $0.2~r_{500}-r_{500}$ and $1.2$ keV in the region $r_{500}-r_{200}$ if clumping correction is neglected. 
\subsection{Importance of non-thermal pressure}
Although, the inclusion of the non-thermal pressure decreases the feedback profiles up to $r_{500}$ due to the overall increase in the normalization ($K_{200}$) in the benchmark entropy profiles, however, as seen from Fig.~\ref{fig:average}, it unexpectedly increases beyond that.  The cross-over occurs around $(1.1-1.2)~r_{500}$. This can be understood as follows: Due to the neglect of the non-pressure, the $M_{tot}$ profile is underestimated which in turn results in the under-estimation of theoretical gas mass as $f_{g,th}=0.9f_b$ is fixed at the virial radius.  This implies for the given observed gas mass shell the corresponding theoretical gas mass  for $P_{nt}=0$ case will occur at larger radius  as compared to $P_{nt}\neq0$ case and  will, therefore, have higher theoretical entropy leading to decrease in feedback profiles. Below $r_{500}$, the increase in $K_{200}$ term dominates (since non-thermal pressure is small) and therefore, there is an overall decrease in the feedback profiles for $P_{nt}\neq0$ case.

From Fig.~\ref{fig:average}, it is evident that ignoring the non-thermal pressure leads to an over-estimation of $\Delta K\approx 100$ keV cm$^2$ at $r_{500}$ and under-estimation of $\Delta K\approx 450$ keV cm$^2$ at $r_{200}$. This in turn leads to an  over-estimation of  $\Delta E \approx 0.5$ keV at $r_{500}$ and under-estimation of $\Delta E\approx 0.25$ keV  at $r_{200}$. Similarly, one can see from Tabs.~\ref{amrplanck}~\&~\ref{amrwmap},  if non-thermal pressure is ignored then the  $\epsilon_{feedback}$ is over-estimated by $0.6$ keV in the region $0.2~r_{500}-r_{500}$ while it has a negligible affect in the region $r_{500}-r_{200}$.

In Fig.~\ref{nonthermalnor}, we show the effect on the feedback profile by changing the normalization $a_0$ and slope $n_{nt}$ in the non-thermal pressure. We find that changing the normalization from 0.18 to (0.10, 0.26) gives around (10\%, 30\%) mass difference at $r_{500}$. The change in normalization and slope has a small effect on the profiles and our results are still consistent with zero line given the error bars. We also observe that other parameterization of non-thermal pressure such as that of \cite{Shi2015,Rasia2004}  lie  within the normalization $0.10-0.26$ band of our non-thermal pressure model. Therefore, our results are independent of the non-thermal parameterization.
\begin{figure*}
\begin{minipage}{8.5cm}
 \includegraphics[width = 8.5cm]{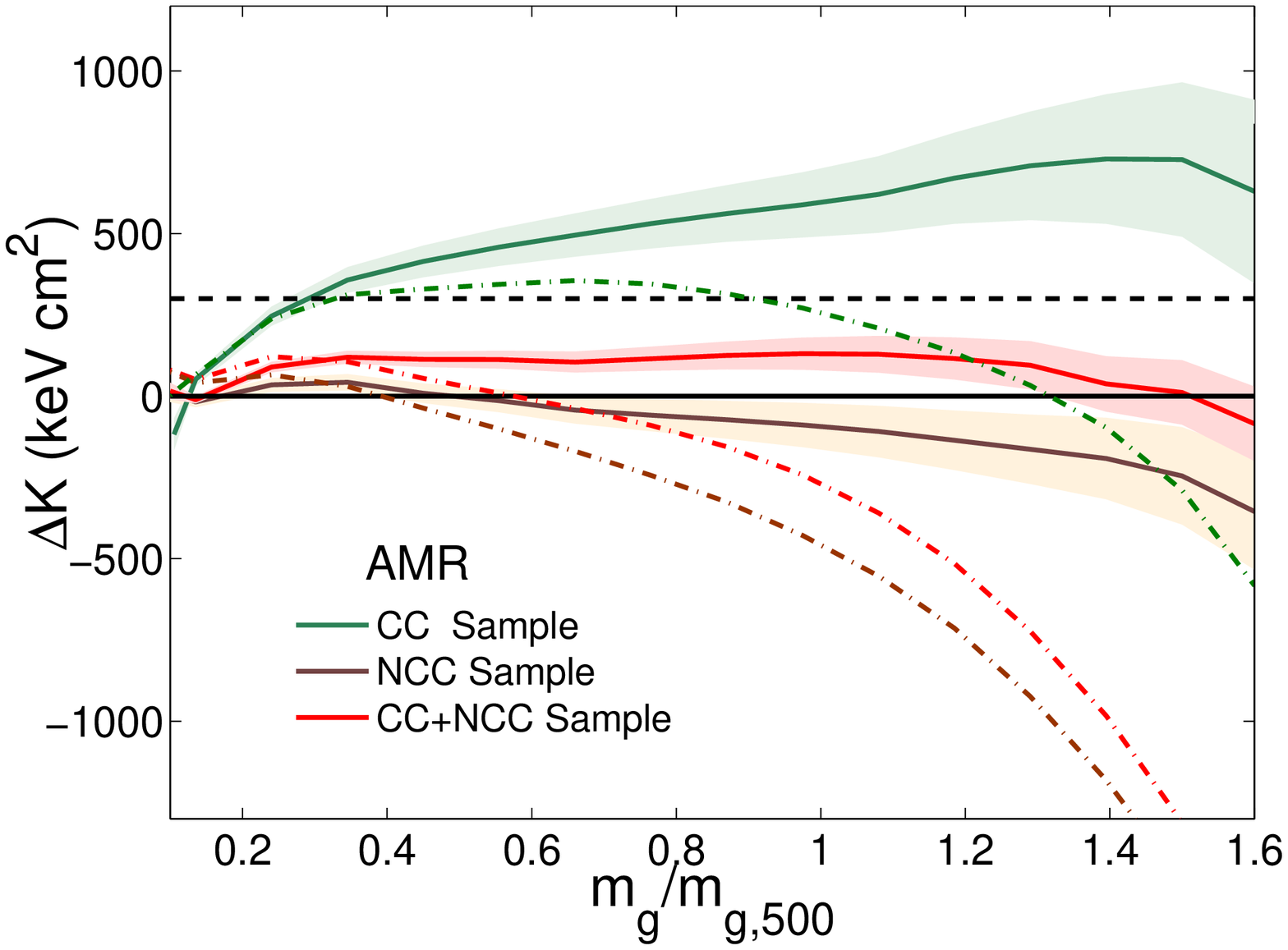}
\end{minipage}
\begin{minipage}{8.5cm}
 \includegraphics[width = 8.5cm]{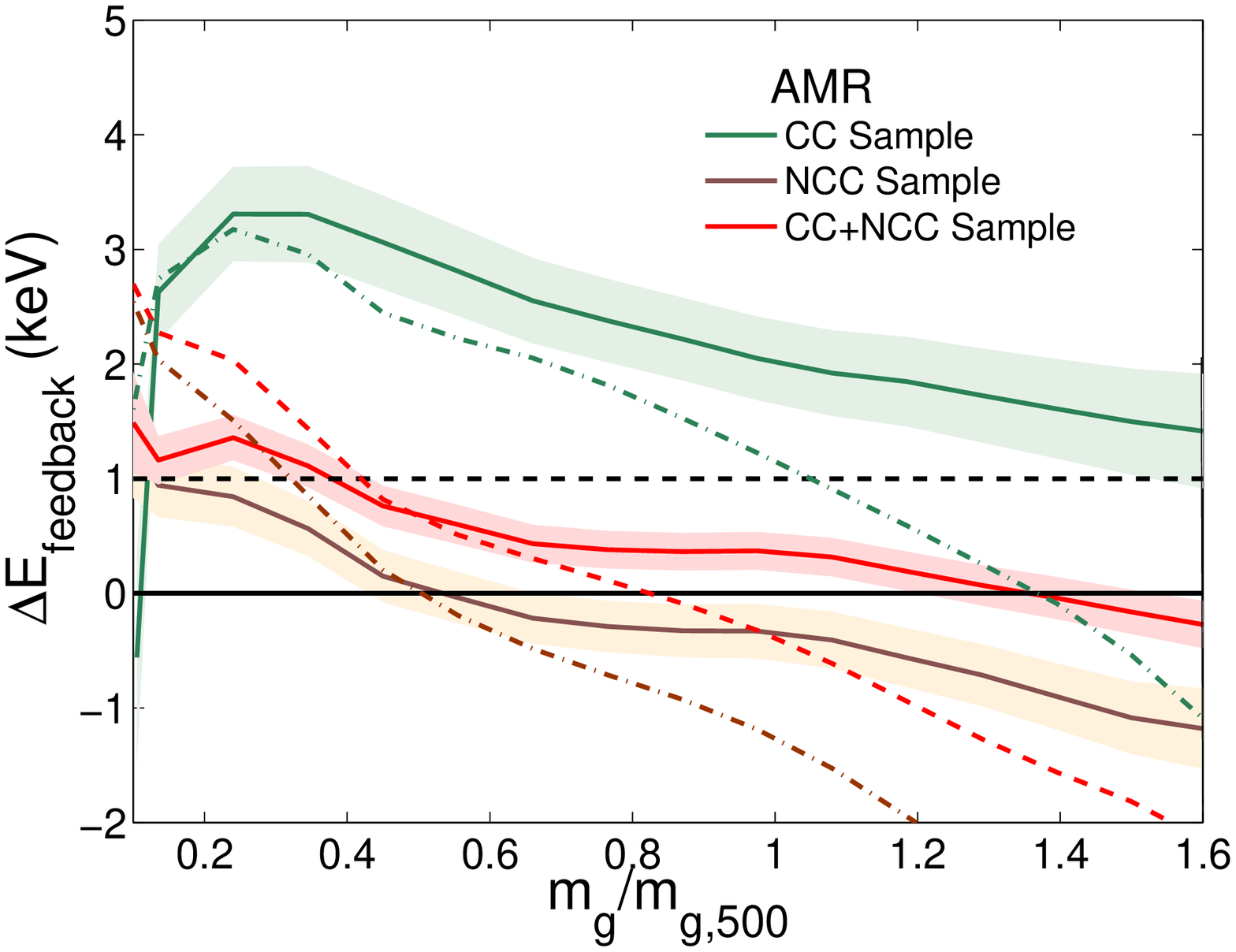}
\end{minipage}   
\caption{Comparison between CC clusters and NCC clusters. The solid lines with error bars represent feedback profiles with  $P_{nt}$ and clumping included and dashed dotted lines (without error bars) represents a case where we do not take  $P_{nt}$ and clumping into account. The error bars are given at 1$\sigma$ level.
}
\label{fig:CCandNCC}
\end{figure*}
\section{Robustness of Results}
\subsection{CC and NCC clusters}
Fig~\ref{fig:CCandNCC} shows the $\Delta K$ and $\Delta E_{feedback}$ profiles for CC  and NCC populations with and without taking into account $P_{nt}$+clumping. The higher value of $\Delta K$ and $\Delta E_{feedback}$ profiles for CC clusters can be interpreted in terms of the gas mass fraction. \cite{Eckert2013b}  found that the observed gas mass fraction profile in the CC clusters was systematically lower than NCC clusters. Since, the theoretical value $f_{g,th}$ is fixed at $0.9f_b$ at virial radius and the observed value of $f_{g,obs}$ for CC clusters is relatively smaller than NCC clusters, this means that for a given observed gas mass $m_{g,obs}$, the corresponding theoretical gas $m_{g,th}$ for CC clusters will also occur at a relatively smaller radius with a smaller value of theoretical entropy. This will thus result in a relatively higher degree of feedback and up to a much larger radii in CC clusters compared to NCC clusters. 

Higher estimates of the feedback profiles in the CC clusters can be due to larger rate of gas removal as a result cooling and simultaneous  inflow of high entropy gas to the cluster cores  (a sort of ``cooling flow''). Moreover, removal of gas due to cooling that corresponds to stellar formation also changes the mapping of the observed gas mass enclosed with radius to the mass shell in the theoretical prediction. This could again contribute to the apparent entropy excess in CC clusters relative to NCC clusters especially in the cluster cores. However, the replacement of cool gas with gas at high entopy occurs mostly in the inner region (few hundred Kpc) which we have not considered in our analysis, and is unlikely to cause much difference.

Another potential origin of the apparent entropy excess in CC clusters could be due to the differences in the shape of the non-thermal pressure profile. Since cool-core clusters tend to have more relaxed centers than NCC clusters, steeper gradient would lead to an over-estimation of total mass and hence a smaller value of observed gas fraction. However, this should compensate  with the higher gas density of CC clusters in the inner regions. Moreover, as can be seen from Fig~\ref{nonthermalnor} the change in normalization and slope of the non-thermal pressure has a small effect in the feedback profiles suggesting this is not the case.
\begin{figure*}
\begin{minipage}{8.5cm}
 \includegraphics[width = 8.5cm]{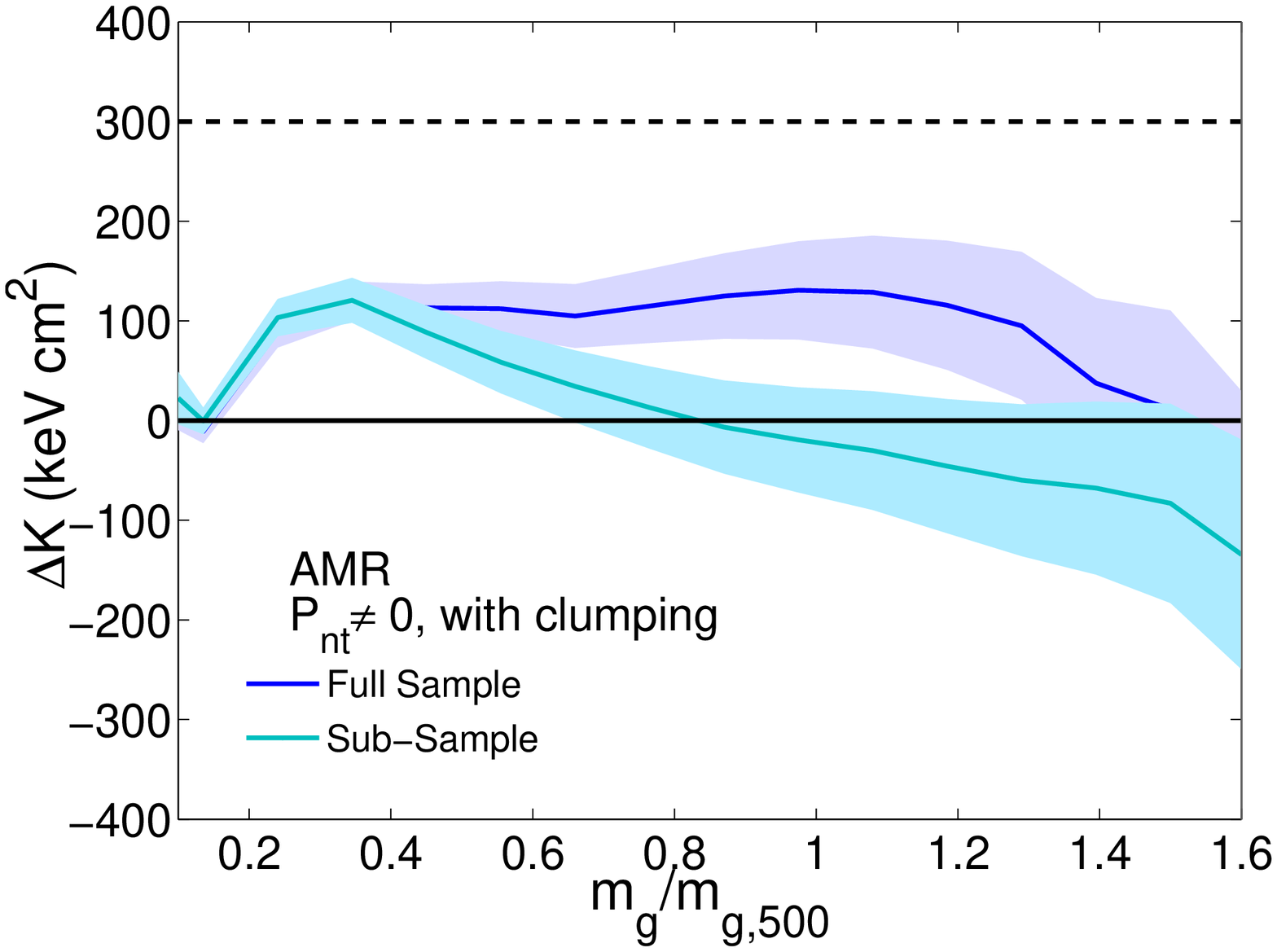}
\end{minipage}
\begin{minipage}{8.5cm}
 \includegraphics[width = 8.5cm]{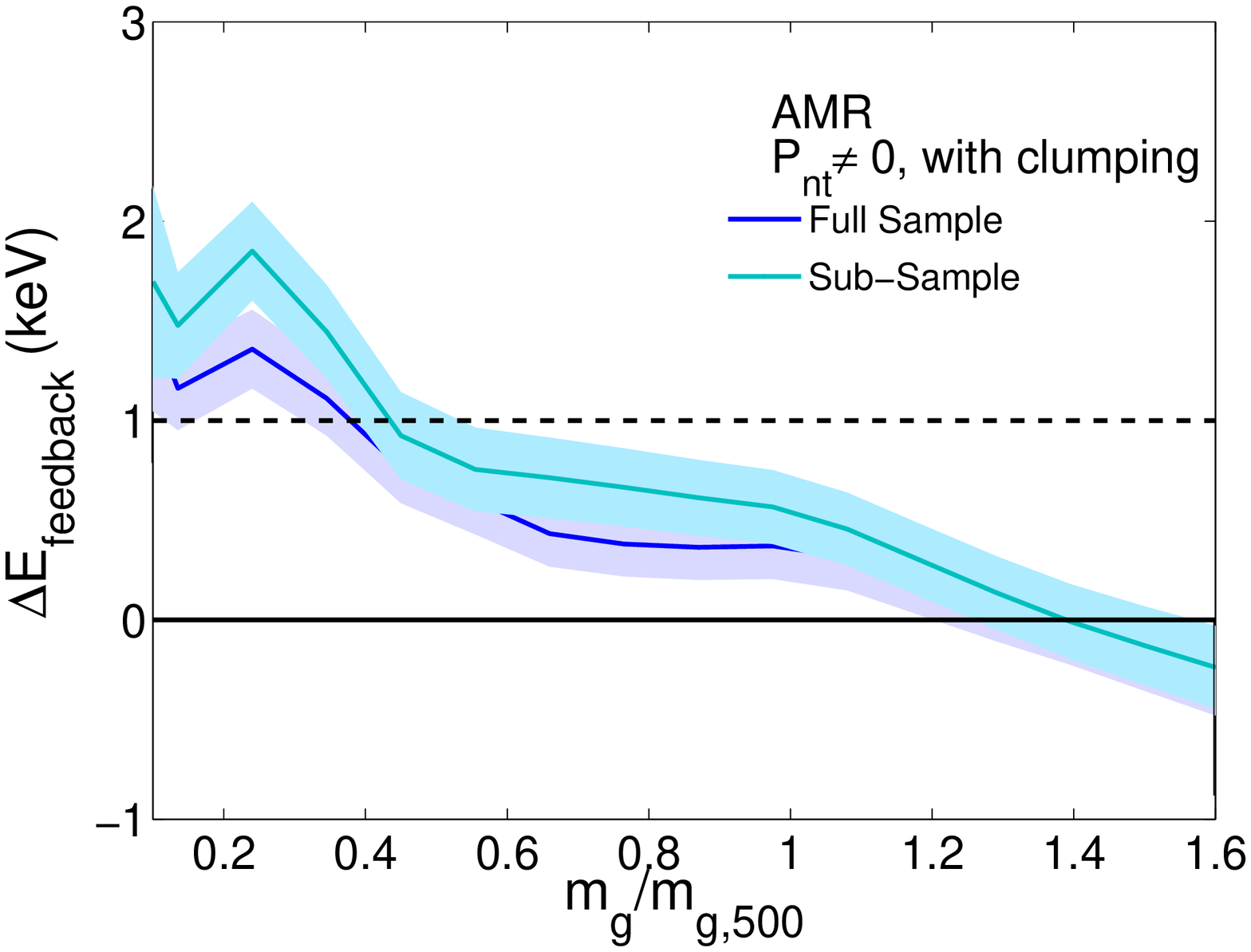}
\end{minipage}   
\caption{Comparison of feedback profiles for sub sample and full sample. The error bars are given at  1$\sigma$ level.}
\label{subsample}
\end{figure*}
\subsection{Full sample and sub-sample}
There are four clusters (i.e clusters 1, 6, 16 and 17) which have relatively large value of the $\Delta K$ profile in the outer regions (particularly for the clumping case) which correspondingly give a large thermal energy profile $\Delta Q_{ICM}$. However, we see that after taking into account  a potential energy term the  $\Delta E_{feedback}$ profiles for such clusters become close to zero (or even negative).
In order to see the effect of such clusters, we have plotted in Fig.~\ref{subsample} average $\Delta K$ and $\Delta E_{feedback}$ profiles for the full sample along with the sub sample which do not include theses clusters. We find that the average feedback entropy become consistent with $\Delta K\approx0$ line at 1$\sigma$ beyond $r_{500}$ for the sub-sample.  Moreover, average $\Delta K$ and $\Delta E_{feedback}$ profiles for the sub-sample and full sample are  always consistent with one another.
\subsection{Choice of boundary condition}
\begin{figure}
\begin{minipage}{8.5cm}
\includegraphics[width=8.5cm]{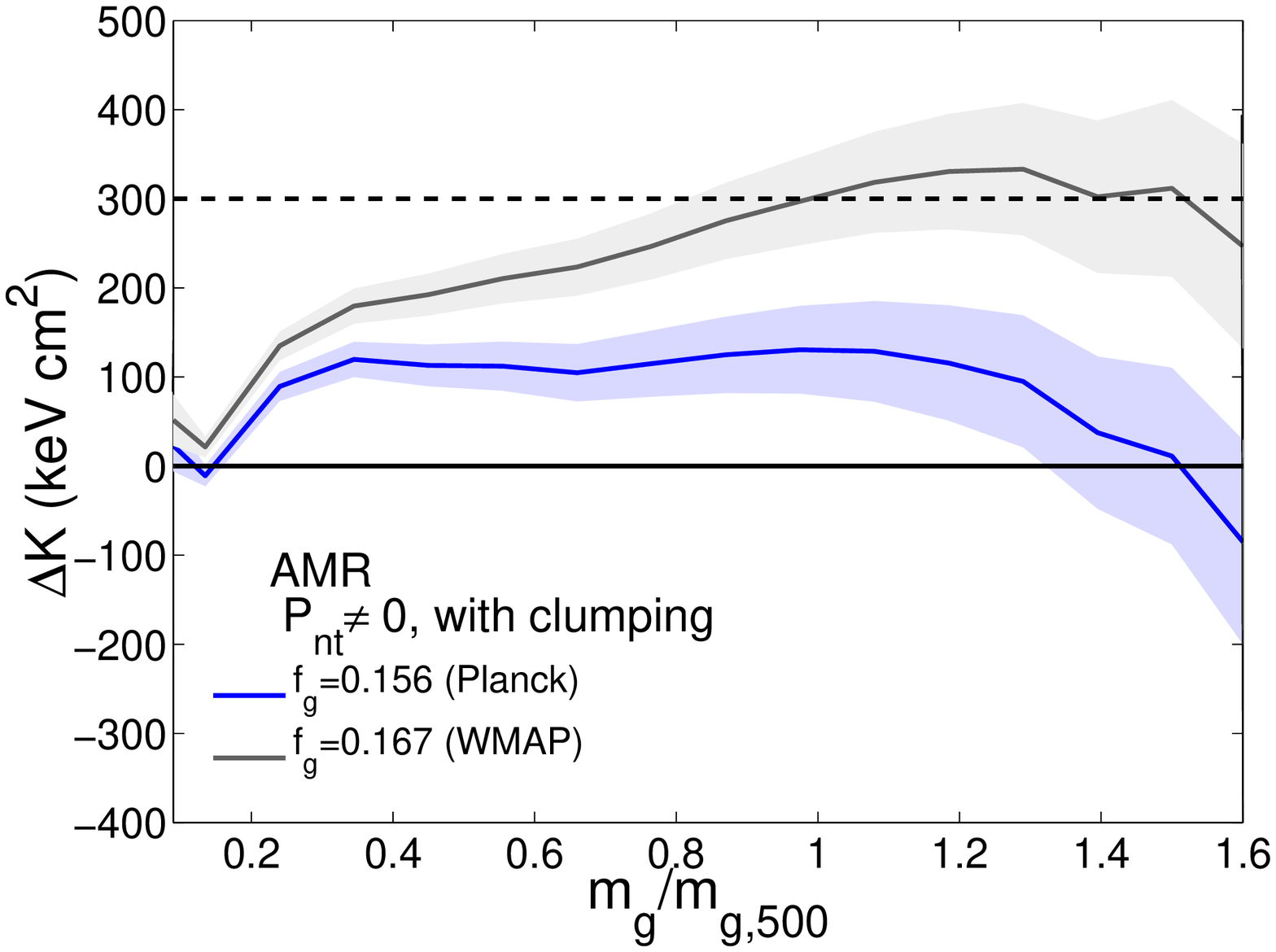}
\end{minipage}
\caption{The average $\Delta K$ feedback profiles as a function of  $m_g/m_{g,500}$ for two boundary conditions of gas fraction at virial radius (i.e $f_{b}=0.156$ from Planck and $f_{b}=0.167$ from WMAP).}
\label{fig:Deprojected}
\end{figure}

In Fig.~\ref{fig:Deprojected}, we show the comparison of the entropy feedback profiles for two different boundary conditions, i.e universal baryonic fraction $f_b$ at virial radius to be $0.156$ (from Planck) and $0.167$ (from WMAP). The larger value of the $f_b$
from WMAP, would result in an overall increase of feedback profiles. This is because higher value of $f_b$ at the virial radius will increase the total theoretical gas mass profile (as total mass remains constant). Therefore, a given theoretical gas mass shell would occur at a smaller radius having a lower value of entropy leading to an increase in feedback profiles. It is clear from Fig.~\ref{fig:Deprojected}, that the  entropy feedback profiles for the WMAP boundary conditions is significantly higher at the outer regions ($\Delta K\approx300$ keV cm$^2$).
\subsection{Choice of observed X-ray profiles - parametric vs deprojected}
\begin{figure}
\begin{minipage}{8.5cm}
\includegraphics[width=8.5cm]{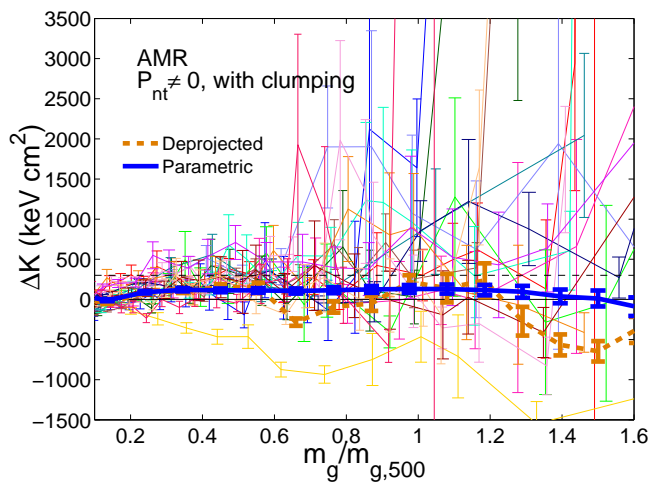}
\end{minipage}
\caption{The average $\Delta K$ feedback profiles as a function  of $m_g/m_{g,500}$ for all the clusters considering the deprojected case.}
\label{fig:withfg}
\end{figure}

\cite{Eckert2013a}  found that the parametric and deprojected density profiles are  similar and the difference is less than 10\%. We show the $\Delta K$ profiles for all the clusters using deprojected data in Fig.~\ref{fig:withfg}. We find parametric and deprojected profiles have similar values of entropy difference from the base theoretical entropy profiles except at the cluster outskirts.   



\begin{figure*}
\begin{minipage}{8.5cm}
 \includegraphics[width = 8.5cm]{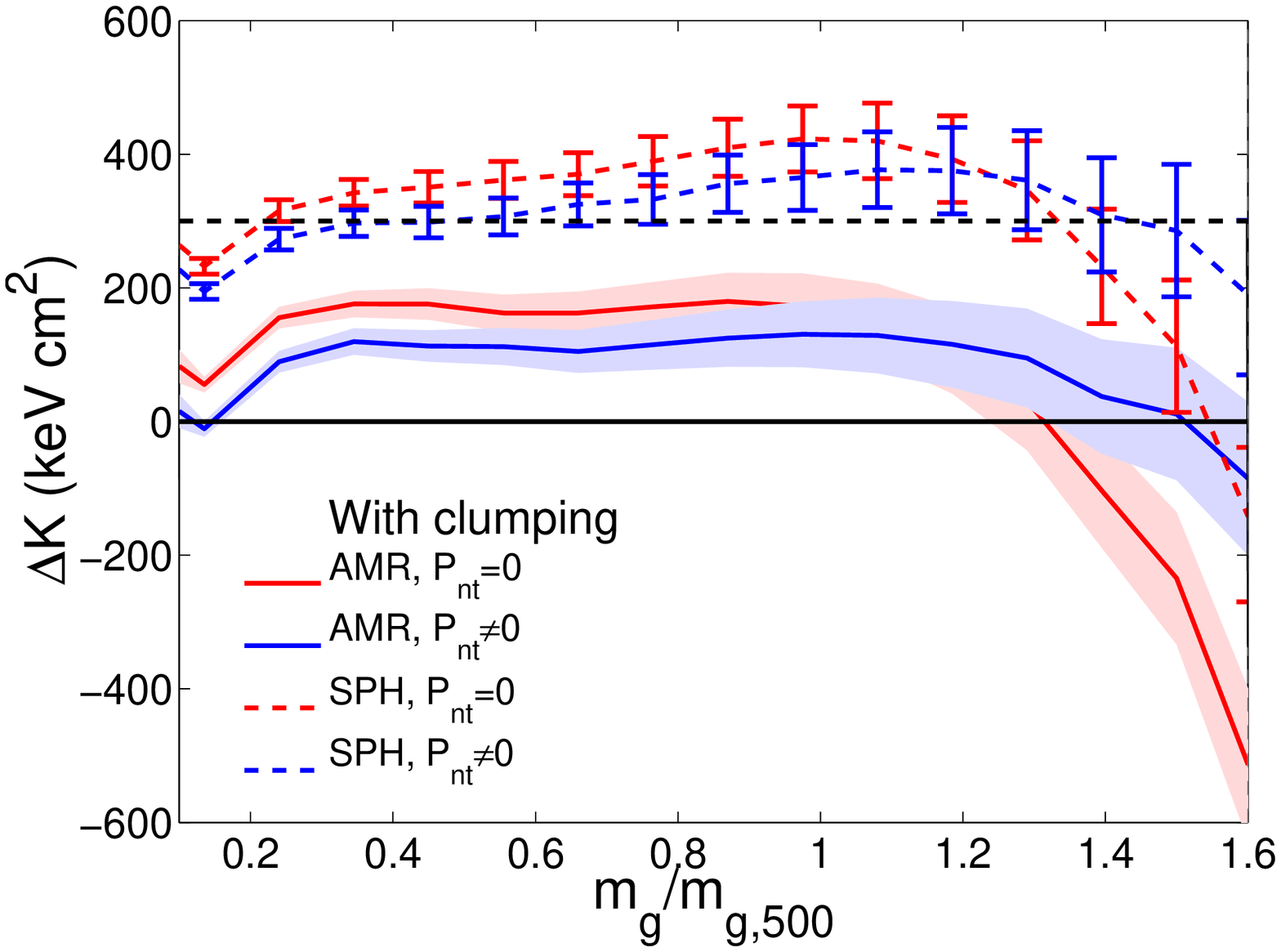}
\end{minipage}
\begin{minipage}{8.5cm}
 \includegraphics[width = 8.5cm]{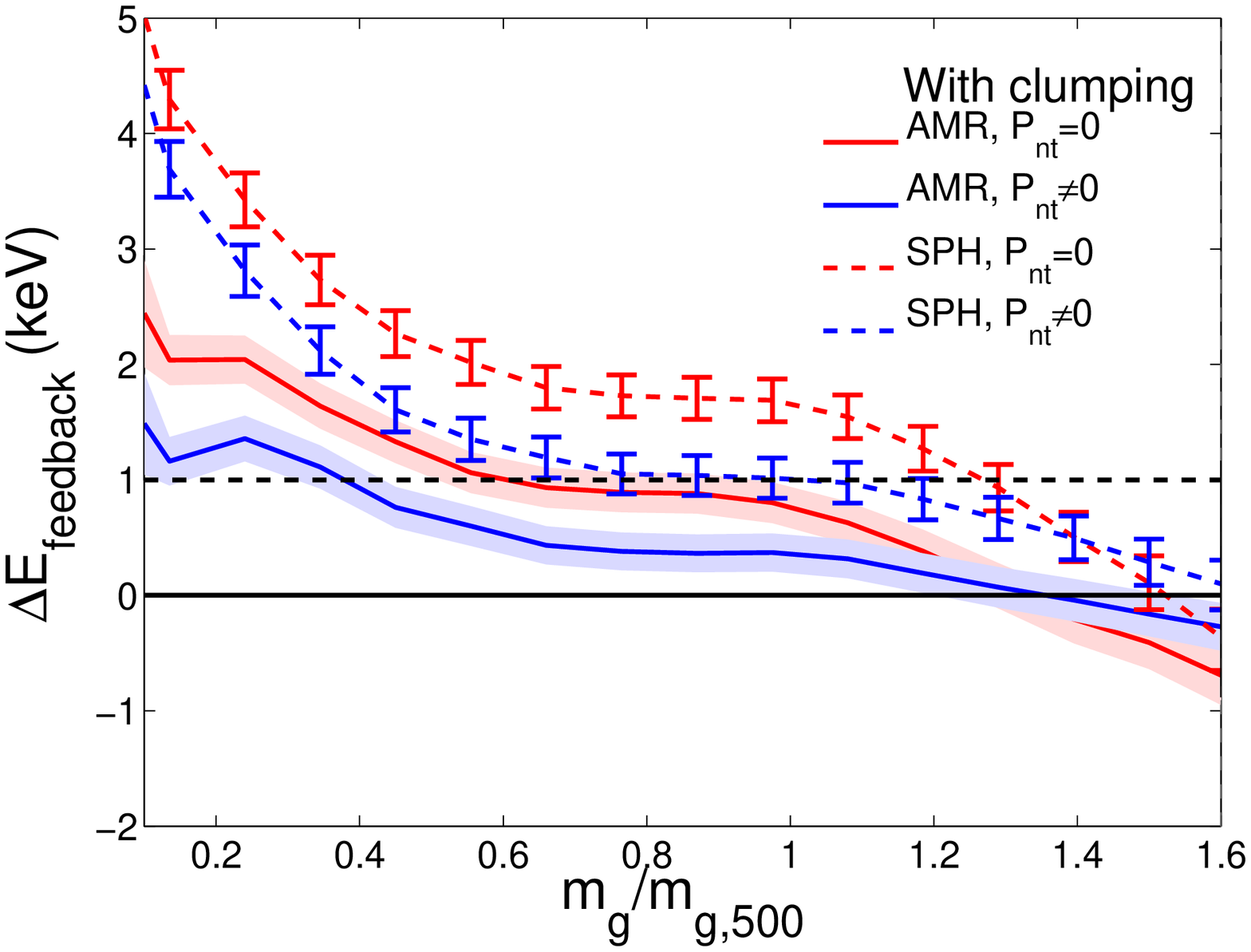}
\end{minipage}   
\caption{Comparison of feedback profiles for AMR and SPH cases. For meaningful comparison, we have scaled x-axis of all  with same $m_{g,500}$ as that of fiducial case (i.e with clumping  and $P_{nt}\neq0$).}
\label{AMRSPHcomp}
\end{figure*}

\section{AMR vs SPH benchmark theoretical profile} 
Given the current status, it is difficult to judge whether SPH or AMR is more accurate since  both these methods are known to have some demerits. For example SPH suffers from a relatively poor shock resolution and 
noise on the scale of the smoothing kernel. AMR simulations may suffer from over-mixing
due to advection errors in the presence of bulk flows. Apart from the slightly smaller normalization of the entropy profile $a_0$, the AMR simulations predict a much higher flatter entropy core than the corresponding SPH simulations \citep{Voit2005}.
However, it has been recently pointed out by many authors that, after resolving certain hydrodynamic processes, the results of SPH simulations exactly match with AMR simulations \citep{Mitchell2009,Vazza2011,Valdarnini2012,Power2014,Biffi2015}.
We nevertheless use SPH simulated entropy in our estimates to compare it with that of AMR case. The feedback profiles and feedback energy per particle for SPH case are shown in the Figs.~\ref{SphdeltaK}~\&~\ref{SPH-comp} and Tabs.~\ref{sphplanck}~\&~\ref{sphwmap} respectively in the appendix \ref{sec:appendix}. In Fig.~\ref{AMRSPHcomp}, we compare the average excess entropy and energy profiles for both AMR and SPH cases.
As can be seen, SPH case predict much higher values of $\Delta E$ in the inner regions of the cluster because of the absence of flatter entropy profile as is present in AMR case. 

The $\Delta K$ and $\Delta E_{feedback}$ feedback profiles at the outer radii contain information of past events in the cluster and hence are ideal to probe for any signature of pre-heating that may have taken place at high redshifts much before the cluster formation. This is because  of the fact that feedback processes from AGN or supernovae are unlikely to affect the gas properties there. However, we should also point that our estimated $\Delta E_{feedback}$ profiles corresponds to the change from the initial theoretical model to the observed configuration in the collapsed systems. Pre-heating (if any) at hight redshift when the  density of gas was small  would actually require much smaller energy input to bring it to final observed state \citep{McCarthy2008}. Therefore, $\Delta E_{feedback}$ would represent an upper limit on pre-heating energy. 

It has been found that pre-heating scenarios (at $z\approx4-6$) typically require feedback energy of $\sim1$ keV per particle or an  entropy floor of $>300$ keV cm$^2$  to explain break in the self-similarity scaling relations \citep{Borgani2001,Tozzi2001,Pipino2002,Finoguenov2003}.
Our results show that given the uncertainties, the values of $\Delta E$ at the outer radii are comparable to zero for both AMR and SPH cases (see Fig.~\ref{AMRSPHcomp}). For our fiducial case, we see that in the range $r_{500}-r_{200}$, the average energy per particle $\epsilon_{feedback}=0.05\pm0.18$ for the AMR case and $\epsilon_{feedback}=0.62\pm0.18$ for SPH case.
This implies that pre-heating scenarios  which predict 1 keV energy per particle are ruled out with more than 3$\sigma$ for AMR case and at around 2$\sigma$ for SPH case. Considering $\Delta K$ profiles, we find for most of the cluster region an entropy floor $>300$ keV cm$^2$ is ruled out at $\approx 3\sigma$ for the AMR case. However, no such strong constrains are possible for the SPH case and that $\Delta K\approx300$ keV cm$^2$ is consistent with $1\sigma$ as seen in the left side of Fig. \ref{AMRSPHcomp}. 
\begin{table*}
 \caption{Average feedback energy per particle $\epsilon_{feedback}$ for AMR case with Planck $f_{b}=0.156$.}
 \label{amrplanck}
 \begin{tabular}{@{}lccccccc}
  \hline
 &&\multicolumn{6}{c}{Energy per particle (keV) }\\
    \cline{3-8}
&&\multicolumn{3}{c}{Without cooling energy}&\multicolumn{3}{c}{With cooling energy}\\
\cline{3-8}
 $C$    &$P_{nt}$   & $0.2-1~r_{500}$ &   $0.2-1~r_{200}$& $r_{500}-r_{200}$&   $0.2-1~r_{500}$&$0.2-1~r_{200}$&$r_{500}-r_{200}$\\
  \hline
\multicolumn{8}{c}{Full Sample}\\
\hline
 0            &0          &$0.39\pm0.20$    &$-0.29\pm0.21$  &  $-1.33\pm0.23$  & $0.80\pm0.20$     &$-0.02\pm0.21$   & $-1.27\pm0.23$    \\
 0            & nonzero   & $-0.31\pm0.19$  &$-0.73\pm0.20$  &  $-1.35\pm0.21$  & $ 0.09\pm0.19$      &$-0.46\pm0.20$   &  $-1.29\pm0.21$     \\
nonzero       & 0         &$0.91\pm0.18$    &$0.60\pm0.19$  &  $0.06\pm0.20$  & $ 1.29\pm0.18$     &$0.85\pm0.19$   & $ 0.11\pm0.20$     \\
nonzero       & nonzero   & $ 0.35\pm0.17$  &$0.23\pm0.17$  &  $0.03\pm0.18$  & $0.72\pm0.17$      &$0.46\pm0.17$   &  $ 0.05\pm0.18$    \\
  \hline 
\multicolumn{8}{c}{NCC Clusters}\\
\hline
 0       &0          &$-0.25\pm0.24$    & $-1.08\pm0.26$  &  $-2.35\pm0.30$  &$0.16\pm0.24$      &$-0.80\pm0.26$    & $-2.28\pm0.30$  \\
 0       & nonzero   &$-0.97\pm0.23$   &$-1.51\pm0.25$   & $-2.35\pm0.27$   & $-0.56\pm0.23$    &$-1.24\pm0.25$    &  $-2.28\pm0.27$     \\
nonzero  & 0         &$0.27\pm0.24$    & $-0.15\pm0.27$  &  $-0.88\pm0.30$  &$0.65\pm0.24$      &$0.09\pm0.27$    & $-0.82\pm0.30$     \\
nonzero  & nonzero   &$-0.21\pm0.23$   &$-0.42\pm0.25$   & $-0.76\pm0.27$   & $0.15\pm0.23$    &$-0.17\pm0.25$    &  $-0.71\pm0.27$    \\
  \hline 
\multicolumn{8}{c}{CC Clusters}\\
\hline
 0        &0          &$1.73\pm0.38$    & $1.06\pm0.38$  &  $0.11\pm0.36$  & $2.14\pm0.38$  &$1.32\pm0.38$    & $  0.16\pm0.36$       \\
 0        & nonzero   & $1.03\pm0.36$   &$0.63\pm0.35$   & $0.12\pm0.33$   & $1.44\pm0.36$  &$ 0.89\pm0.35$    &   $ 0.17\pm0.33$     \\
nonzero   & 0         &$2.52\pm0.39$    & $2.36\pm0.41$  &  $2.02\pm0.44$  & $2.88\pm0.39$  &$2.60\pm0.41$    & $ 2.06\pm0.44$     \\
nonzero   & nonzero   & $2.20\pm0.41$   &$1.97\pm0.41$   & $1.68\pm0.34$   & $2.60\pm0.41$  &$2.23\pm0.41$    &   $1.73\pm0.41$    \\
  \hline 

\end{tabular}

 Columns (3), (4) \& (5): $\epsilon_{feedback}$ in the ranges $(0.2-1)~r_{500}$, $(0.2-1)~r_{200}$ and $r_{500}-r_{200}$ respectively without taking into account energy lost due to cooling. 
 Columns (6), (7) \& (8): $\epsilon_{feedback}$ in the ranges $(0.2-1)~r_{500}$, $(0.2-1)~r_{200}$ and $r_{500}-r_{200}$ respectively after taking into account energy lost due to cooling. The errors are given at 1$\sigma$ level.
For meaningful comparison, $\epsilon_{feedback}$ for $P_{nt}=0$ case are also calculated up to same radii as that of non-thermal case (i.e $r_{500}$ and $r_{200}$ of $P_{nt}\neq0$).
\end{table*}

\begin{table*}
 \caption{Average feedback energy per particle $\epsilon_{feedback}$ for AMR case with WMAP $f_{b}=0.167$.}
 \label{amrwmap}
 \begin{tabular}{@{}lccccccc}
  \hline
 &&\multicolumn{6}{c}{Energy per particle (keV) }\\
    \cline{3-8}
&&\multicolumn{3}{c}{Without cooling energy}&\multicolumn{3}{c}{With cooling energy}\\
\cline{3-8}
 $C$    &$P_{nt}$   & $0.2-1~r_{500}$ &   $0.2-1~r_{200}$& $r_{500}-r_{200}$&   $0.2-1~r_{500}$&$0.2-1~r_{200}$&$r_{500}-r_{200}$\\
  \hline
\multicolumn{8}{c}{Full Sample}\\
\hline
 0        &0          &$0.89\pm0.20$    &$ 0.22\pm0.21$  &  $-0.79\pm0.21$  & $1.30\pm0.20$     &$0.49\pm0.21$   & $-0.72\pm0.21$\\
 0        & nonzero   & $0.17\pm0.19$   &$-0.22\pm0.19$  & $-0.83\pm0.20$   & $0.58\pm0.19$    &$0.04\pm0.19$   &  $-0.77\pm0.20$       \\
nonzero   & 0         &$1.35\pm0.19$    &$1.11\pm0.19$   & $ 0.71\pm0.19$    &$1.73\pm0.19$      &$1.37\pm0.19$   &$0.76\pm0.19$\\
nonzero   & nonzero   & $0.72\pm0.18$   &$ 0.66\pm0.17$   &  $0.56\pm0.17$   &$1.10\pm0.18$      &$0.89\pm0.17$   &$0.59\pm0.17$\\
  \hline 
\multicolumn{8}{c}{NCC Clusters}\\
\hline
 0         &0          &$0.26\pm0.24$    & $-0.56\pm0.26$  & $-1.83\pm0.26$ &$ 0.68\pm0.26$     &$-0.28\pm0.26$ &$-1.70\pm0.29$\\
 0         & nonzero   &$-0.45\pm0.23$  &$-0.99\pm0.24$  &$-1.83\pm0.25$  & $-0.04\pm0.23$   &$-0.72\pm0.24$&  $-1.76\pm0.25$       \\
nonzero    & 0         &$0.73\pm0.25$    &$0.41\pm0.26$   & $-0.12\pm0.29$  &$1.11\pm0.25$    &$0.67\pm0.26$ & $-0.07\pm0.29$\\
nonzero    & nonzero   & $0.16\pm0.24$  &$0.05\pm0.25$  &$-0.11\pm0.26$  &$0.54\pm0.24$     &$0.30\pm0.25$ &$-0.06\pm0.26$\\
  \hline 
\multicolumn{8}{c}{CC Clusters}\\
\hline
 0        &0          &$2.28\pm0.38$    &$1.69\pm0.38$  &$0.88\pm0.36$  & $2.69\pm0.38$   &$1.94\pm 0.38$ &$ 0.93\pm0.36$\\
 0        & nonzero   & $1.55\pm0.36$     &$1.22\pm0.35$  &$0.77\pm0.32$  & $1.96\pm0.36$  &$1.48\pm0.35$ &  $ 0.82\pm0.32$       \\
nonzero   & 0         &$2.98\pm0.40$     &$2.88\pm0.42$  & $2.68\pm0.45$ &$3.34\pm0.40$   &$3.12\pm0.42$&$2.72\pm0.45$\\
nonzero   & nonzero   & $2.70\pm0.41$     &$2.53\pm0.42$  &  $2.33\pm0.42$&$3.10\pm0.41$     &$2.80\pm0.42$ &$2.38\pm0.42$\\
  \hline 
\end{tabular}

 Columns (3), (4) \& (5): $\epsilon_{feedback}$ in the ranges $(0.2-1)~r_{500}$, $(0.2-1)~r_{200}$ and $r_{500}-r_{200}$ respectively without taking into account energy lost due to cooling. 
 Columns (6), (7) \& (8): $\epsilon_{feedback}$ in the ranges $(0.2-1)~r_{500}$, $(0.2-1)~r_{200}$ and $r_{500}-r_{200}$ respectively after taking into account energy lost due to cooling. The errors are given at 1$\sigma$ level.
For meaningful comparison, $\epsilon_{feedback}$ for $P_{nt}=0$ case are also calculated up to same radii as that of non-thermal case (i.e $r_{500}$ and $r_{200}$ of $P_{nt}\neq0$).
\end{table*}
\section{Comparison with previous results}
\begin{figure*}
\begin{minipage}{8.5cm}
\includegraphics[width=8.5 cm]{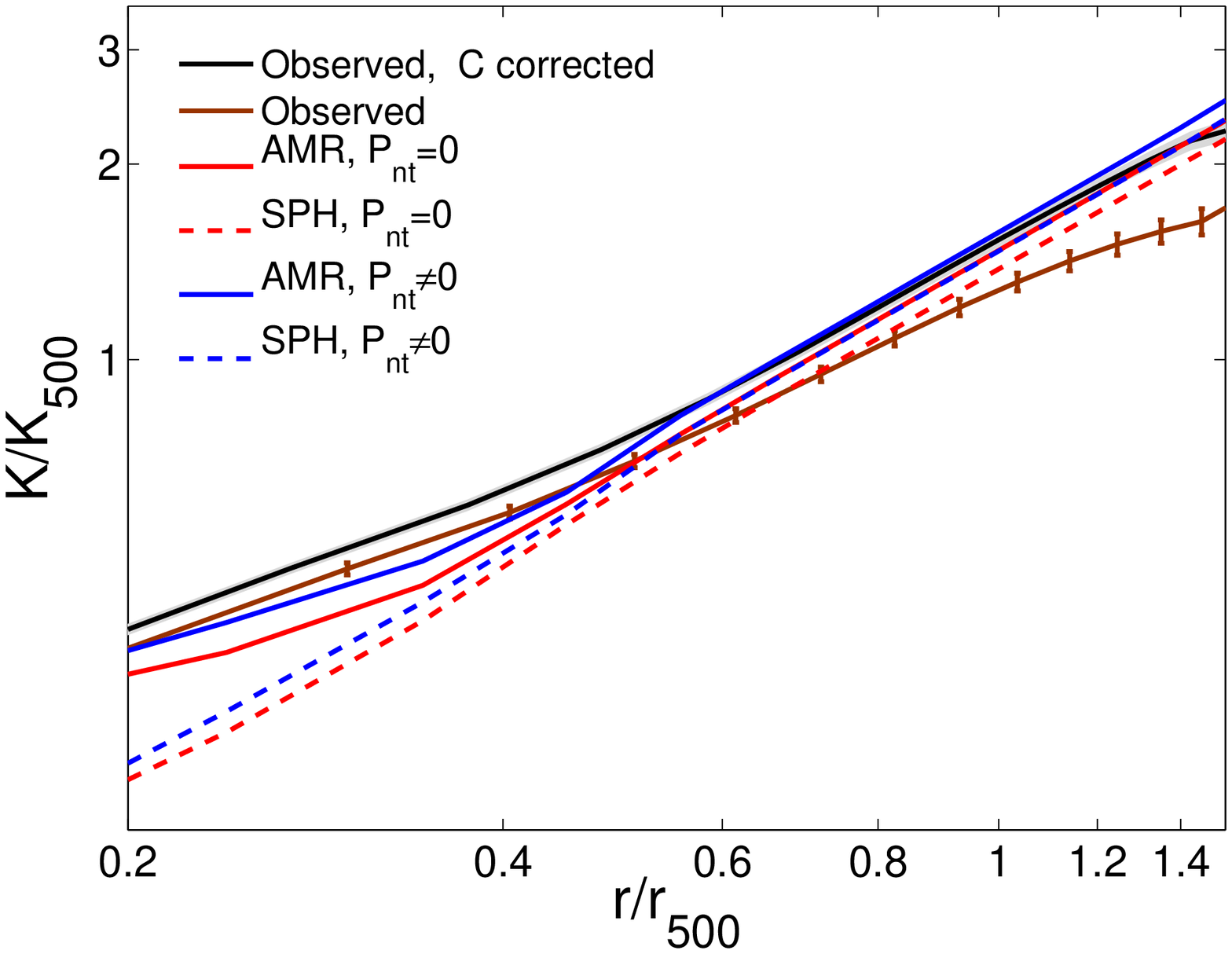}
\end{minipage}
\begin{minipage}{8.5cm}
\includegraphics[width=8.5 cm]{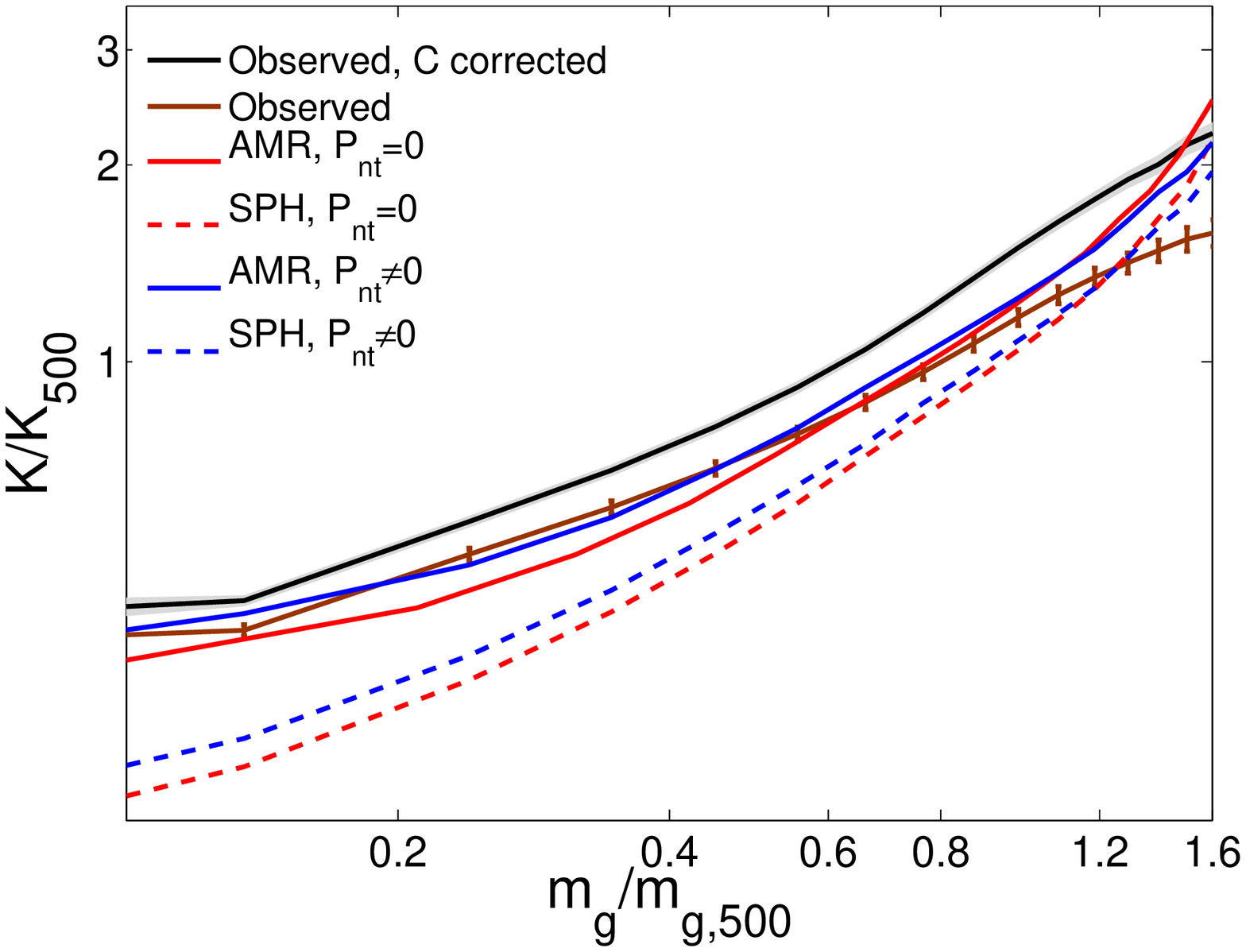}
\end{minipage}
\caption{
Comparison between entropy profiles with respect to fixed radii (left panel) and with respect to shells with a given gas (right panel). The observed average entropy profiles with and without clumping correction are  shown in black and brown lines, whereas theoretical profiles with and without non-thermal pressure are shown in  blue and  red lines respectively. Note that for meaningful comparison, we have scaled $K$, $m_g$ and $r$ with same $K_{500}$, $m_{g,500}$ and $r_{500}$ as that of fiducial case (i.e with clumping  and $P_{nt}\neq0$).}
\label{fig:comp}
\end{figure*}

It is important to distinguish between the profiles of $K$  with respect to shells at fixed position \citep{10,Ettori2013}, and with respect to shells with a given gas mass interior to it \citep{Nath2011,Chaudhuri2012,Chaudhuri2013}. This aspect is demonstrated in Fig. \ref{fig:comp}. The right panel shows the entropy profiles with respect to fixed radii, and the left panel shows the entropy with respect to shells with a given gas mass interior to it. The differences between the profiles (w.r.t. $r/r_{500}$ and $m_g/m_{g,500}$) are striking and noteworthy. The observed entropy profiles (solid brown line) shows an enhancement of entropy in the inner region, but drops below the theoretical profile in the outer region. In this case, the inclusion of non-thermal pressure would seem  to exacerbate the situation and the deviation of observed profile becomes acute. However, one see that after accounting for the clumping correction, the recovered  entropy profiles (solid black line) shows an excess compared to theoretical entropy profile with respect to gas mass for most of the cluster region. Comparing entropy profiles at the same radii, we find the deviations between theoretical and clumping corrected entropy profiles become negligible beyond $0.5~r_{500}$ . It is also worth mentioning here that for theoretical entropy profiles the cross-over between $P_{nt}=0$ and $P_{nt}\neq0$  cases is around $(1.1-1.2)~r_{500}$ as also seen in average $\Delta K$ and $\Delta E$ profiles (see Fig.~\ref{fig:average}).  

Earlier, \cite{Chaudhuri2013} determined the feedback profile up to $r_{500}$.  They found total feedback energy $E_{feedback}$
scales with the mean spectroscopic temperature as $E_{feedback} \propto T^{2.52\pm0.08}$ and $E_{feedback} \propto T^{2.17\pm0.11}$ for the
SPH and AMR baseline profiles respectively. They showed that $E_{feedback}$ correlates strongly with the  radio luminosity $L_R$ of the central radio sources and estimated energy per particle to be $2.8 \pm 0.8$ keV for the SPH simulations and $1.7 \pm 0.9$ keV for the AMR simulations which is much greater than our estimate. 
Notice that they did not consider non-thermal pressure and clumping and their calculations included cluster cores which results in the higher estimates of energy per particle. 
\section{Conclusions}
Recent studies have revealed that non-gravitational processes play an important role in  modifying the thermodynamic properties of the ICM. It has also been observed that there is an entropy enhancement in galaxy clusters which is believed to be  a result of the non-gravitational feedback from  active galactic nuclei, radiative cooling, supernovae etc. 
We have studied the fractional entropy enhancement and the corresponding feedback energy in the ICM for a sample of 17 galaxy clusters by comparing the observed entropy profiles with that of AMR and SPH non-radiative simulations. Unlike, previous work by \cite{Chaudhuri2013} where they estimated the feedback energetics up to radius $\lesssim r_{500}$, our study goes up to $r_{200}$. Since around 30\% of the total cluster mass (and almost 80\% of the cluster volume) is outside of $r_{500}$,
this study has important implication on the  ICM thermodynamics and feedback processes. The cluster outskirts have many features which are not significant into cluster cores. These include, deviation from the hydrostatic equilibrium and gas clumping.  Therefore,  our analysis takes both non-thermal pressure and clumping into account which are important to study the energetics of ICM in the outer regions.  

We show that neglect of  clumping and non-thermal pressure can lead to an under-estimation of $r_{500}$ and $r_{200}$ by $10\%-20\%$. Similarly, we find an under/over-estimation of feedback profiles.  We find that the effect of clumping is much more pronounced than the non-thermal pressure and neglecting it always leads to an under-estimation of feedback profiles. The neglect of clumping leads to an under-estimation of entropy $\Delta K\approx 1100$ keV cm$^2$ and feedback energy per particle $\Delta E_{feedback} \approx 1$ keV at $r_{200}$. The neglect of non-thermal pressure on the other hand lead to an over-estimation in the inner regions and under-estimation in the outer regions. The omission of the non-thermal pressure results in an  under-estimation of entropy $\Delta K\approx 450$ keV cm$^2$ and feedback energy per particle $\Delta E_{feedback}\approx 0.25$ keV  at $r_{200}.$
Further, we  find that the feedback energy profiles are centrally peaked which can be related with AGN feedback models and more or less flatten out in the outer regions becoming consistent with zero.  

Finally, to check the robustness of our results, we compared the feedback profiles for different cases:  CC and NCC clusters, the full sample and a sub-sample, parametric and deprojected cases. We find the much higher value of feedback profiles for the CC clusters compared to NCC clusters. However, since CC clusters comprise  a much smaller sample one, therefore, needs to improve/verify the estimates by considering a larger sample.  We also find that the choice of the universal baryonic fraction from  WMAP and Planck can have implications on the estimates of the feedback profiles.  

In order to obtain any robust estimates of cosmological parameters from large SZ surveys and SZ power spectrum, one needs to  properly incorporate the non-gravitational feedback. Therefore, it is utmost important to understand the nature and extent of the non-gravitational feedback in galaxy clusters, out to the virial radius, so as to properly calibrate the scaling relations and theoretical cluster models. In principle, one can consider different parameterizations of excess entropy and feedback energy to see the affects of the non-gravitational feedback on ICM thermodynamics. 

\section*{Acknowledgements}
This work was supported by SERB (DST) Project Grant No. SR/S2/HEP-29/2012. AI would like to thank Tata Institute of Fundamental Research (TIFR), Mumbai and Raman Research Institute (RRI), Banglore for hospitality.
\footnotesize{
    \bibliographystyle{mn2e}

}
\appendix

\section{SPH results}
\label{sec:appendix}

\begin{figure*}
\begin{minipage}{8.5cm}
 \includegraphics[width = 8.5 cm]{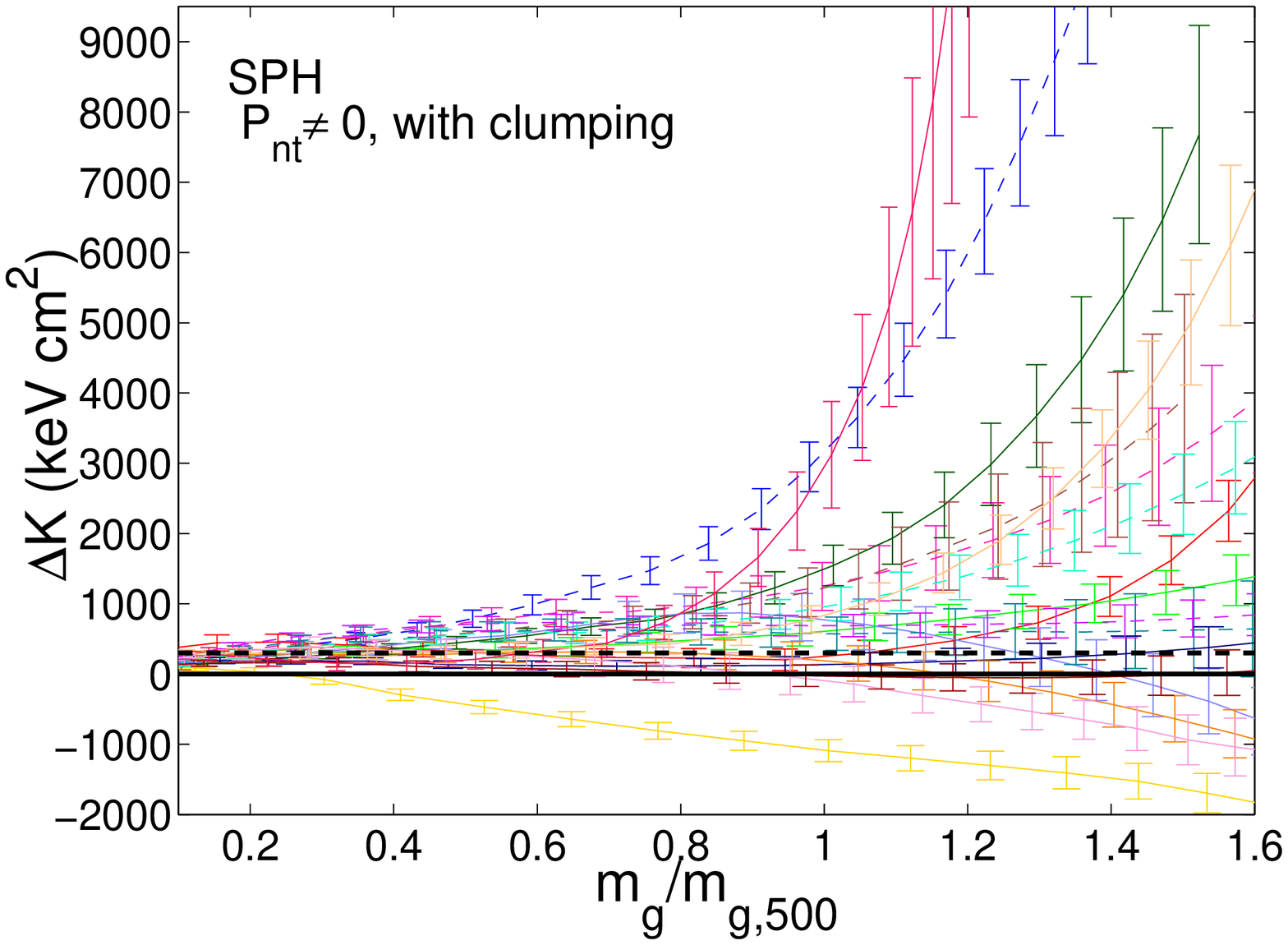}
\end{minipage}
\begin{minipage}{8.5cm}
 \includegraphics[width = 8.5 cm]{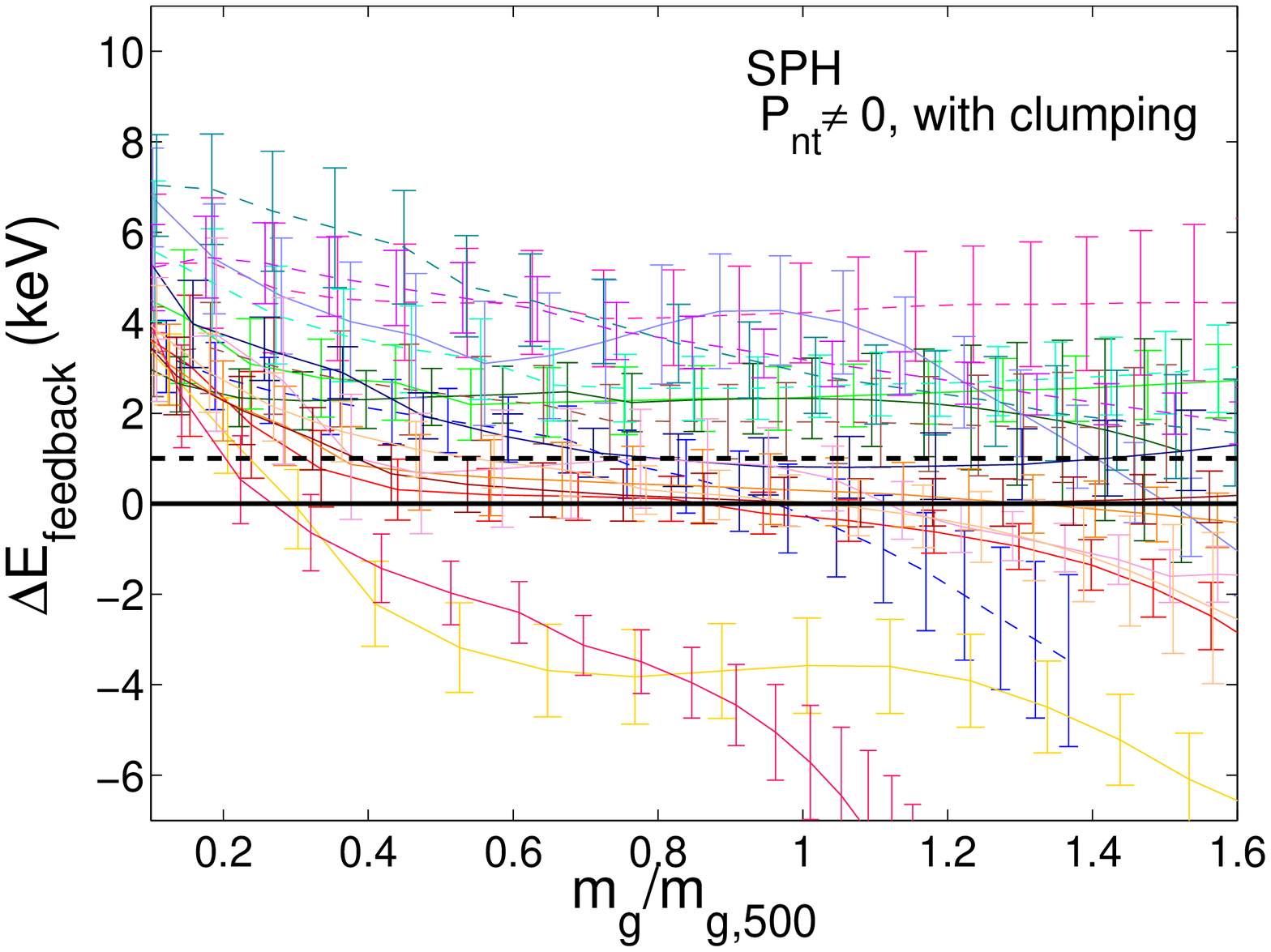}
\end{minipage}   
 \caption{ $\Delta K$ and $\Delta E$ profiles for SPH case.}
\label{SphdeltaK}
 \end{figure*}

\begin{figure*}
\begin{minipage}{8.5cm}
 \includegraphics[width = 8.5 cm]{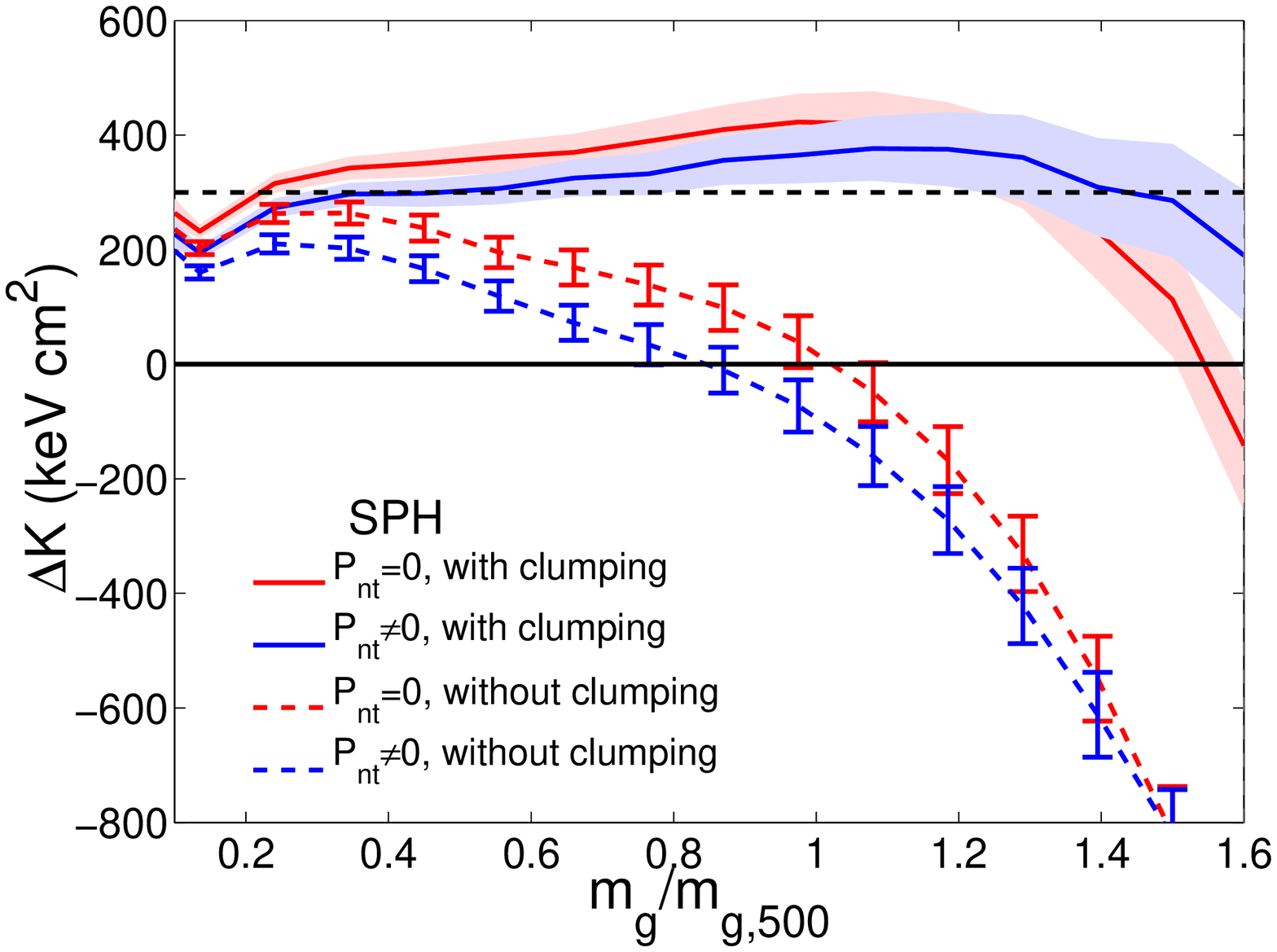}
\end{minipage}
\begin{minipage}{8.5cm}
 \includegraphics[width = 8.5 cm]{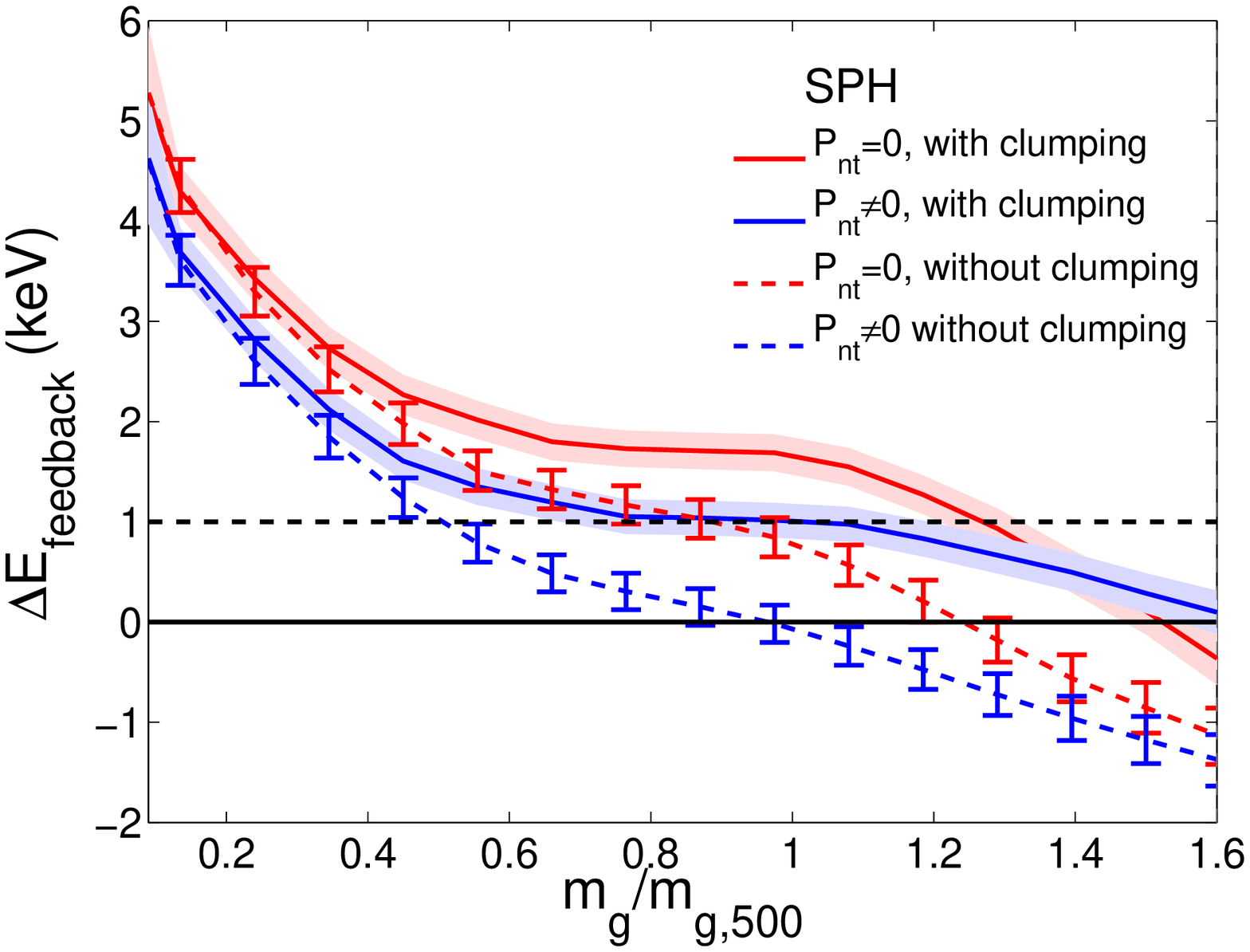}
\end{minipage}   
 \caption{Comparison of $\Delta K$  and $\Delta E$ profiles for SPH case.}
\label{SPH-comp}
 \end{figure*}

\begin{table*}
 \caption{Average feedback energy per particle $\epsilon_{feedback}$ for SPH case with {\it Planck} $f_{b}=0.154$.}
 \label{sphplanck}
 \begin{tabular}{@{}lccccccc}
  \hline
 &&\multicolumn{6}{c}{Energy per particle (keV) }\\
    \cline{3-8}
&&\multicolumn{3}{c}{Without cooling energy}&\multicolumn{3}{c}{With cooling energy}\\
\cline{3-8}
 $C$    &$P_{nt}$   & $0.2-1~r_{500}$ &   $0.2-1~r_{200}$& $r_{500}-r_{200}$&   $0.2-1~r_{500}$&$0.2-1~r_{200}$&$r_{500}-r_{200}$\\
  \hline
\multicolumn{8}{c}{Full Sample}\\
\hline
 0        &0          &$1.61\pm0.21$     &$0.86\pm0.22$  & $-0.25\pm0.22$  &$2.03\pm0.21$      &$1.13\pm0.22$   & $-0.19\pm0.22$     \\
 0        & nonzero   & $0.84\pm0.20$   &$0.19\pm0.20$  & $-0.79\pm0.21$ & $1.25\pm0.20$     &$0.46\pm0.20$   &  $-0.73\pm0.21$    \\
nonzero   & 0         &$2.03\pm0.20$    &$1.59\pm0.20$  &  $0.85\pm0.20$ &$2.41\pm0.20$      &$1.84\pm0.20$   &  $0.90\pm0.20$   \\
nonzero   & nonzero   &$1.39\pm0.19$     &$1.10\pm0.19$  &  $0.61\pm0.18$ &$1.77\pm0.19$     &$1.31\pm0.19$   &  $0.62\pm0.18$   \\\\
  \hline 
\multicolumn{8}{c}{NCC Clusters}\\
\hline
 0        &0          &$ 0.95\pm0.25$    &$0.01\pm0.26$   & $-1.41\pm0.28$   & $1.37\pm0.25$   &$0.28\pm0.26$    &$-1.34\pm0.28$    \\
 0        & nonzero   &$ 0.18\pm0.24$   &$-0.64\pm0.25$  & $-1.89\pm 0.27$   & $0.59\pm0.25$   &$-0.36\pm0.25$    & $-1.82\pm0.26$      \\
nonzero   & 0         &$1.38\pm0.26$  &$0.85\pm0.28$   & $-0.04\pm0.30$    &$1.76\pm0.26$    &$1.10\pm0.28$    &  $0.01\pm0.30$     \\
nonzero   & nonzero   & $0.78\pm0.25$  &$ 0.42\pm 0.26$   & $-0.15\pm0.28$   &$1.16\pm0.25$   &$0.68\pm0.6$    &  $-0.10\pm0.28$     \\
  \hline                                                                     
\multicolumn{8}{c}{CC Clusters}\\
\hline
 0         &0          &$3.21\pm0.41$    & $ 2.57\pm0.40$  &$1.67\pm0.36$   & $3.62\pm0.41$      &$2.83\pm0.40$    &$1.72\pm0.36$     \\
 0         & nonzero   &$2.44\pm0.39$    &$1.85\pm 0.37$   & $0.96\pm0.34$  & $2.85\pm0.39$      &$2.11\pm 0.37$    &$1.01\pm0.34$     \\    
nonzero    & 0         &$3.81\pm0.44$   &$3.50\pm0.45$   & $2.97\pm0.48$ &$4.18\pm0.44$       &$3.74\pm0.45$    &$3.01\pm0.48$     \\
nonzero    & nonzero   &$3.65\pm0.45$   &$3.25\pm0.45$   &$ 2.61\pm0.44$   &$4.05\pm0.45$       &$3.51\pm0.45$    &$2.66\pm0.44$     \\
  \hline 

\end{tabular}
\end{table*}

\begin{table*}
 \caption{Average feedback energy per particle $\epsilon_{feedback}$ for SPH case with WMAP $f_{b}=0.167$.}
 \begin{tabular}{@{}lccccccc}
  \hline
 &&\multicolumn{6}{c}{Energy per particle (keV) }\\
    \cline{3-8}
&&\multicolumn{3}{c}{Without cooling energy}&\multicolumn{3}{c}{With cooling energy}\\
\cline{3-8}
 $C$    &$P_{nt}$   & $0.2-1~r_{500}$ &   $0.2-1~r_{200}$& $r_{500}-r_{200}$&   $0.2-1~r_{500}$&$0.2-1~r_{200}$&$r_{500}-r_{200}$\\
  \hline
\multicolumn{8}{c}{Full Sample}\\
\hline
 0        &0          &$2.02\pm0.21$     &$1.31\pm0.22$   & $0.23\pm0.22$  &$2.43\pm0.21$     &$1.58\pm0.22$  & $0.30\pm0.22$\\
 0        & nonzero   & $1.25\pm0.20$   &$0.64\pm0.20$    & $-0.26\pm0.20$& $1.67\pm0.20$      &$0.91\pm0.20$  &  $-0.20\pm0.20$       \\
nonzero   & 0         &$ 2.38\pm0.20$    &$2.04\pm0.20$    & $1.49\pm0.20$&$ 2.75\pm0.20$       &$2.30\pm0.20$ & $  1.54\pm0.20$ \\
nonzero   & nonzero   & $1.72\pm0.20$   &$1.51\pm0.19$    & $1.17\pm0.18$ &$2.09\pm0.20$      &$1.73\pm0.19$  &$1.18\pm0.18$   \\
  \hline
\multicolumn{8}{c}{NCC Clusters}\\
\hline
 0        &0          &$1.38\pm0.25$    & $0.49\pm0.26$   &$-0.86\pm0.27$& $1.80\pm0.25$&$0.77\pm0.26$ &   $-0.79\pm0.27$ \\
 0        & nonzero   &$ 0.62\pm0.24$   &$-0.15\pm0.25$    & $-1.32\pm0.25$& $1.04\pm0.24$     &$0.12\pm0.25$   & $-1.25\pm0.25$ \\
nonzero   & 0         &$1.74\pm0.27$    &$1.34\pm 0.28$    & $0.69\pm0.30$&$2.12\pm0.27$       &$1.60\pm0.28$ & $0.74\pm0.30$ \\
nonzero   & nonzero   &$1.12\pm0.26$   &$0.89\pm0.26$     & $0.50\pm0.27$&$1.50\pm0.26$      &$1.15\pm0.26$  & $ 0.55\pm0.27$       \\
  \hline 
\multicolumn{8}{c}{CC Clusters}\\
\hline
 0       &0          &$ 3.63\pm0.42$    &$3.08\pm0.41$   &$2.27\pm0.37$ & $4.03\pm0.42$      &$3.34\pm0.41$ &  $2.32\pm0.37$        \\
 0       & nonzero   & $2.87\pm0.40$   &$2.35\pm0.38$    &$1.63\pm0.34$ & $3.28\pm0.40$     &$2.61\pm0.38$   &  $1.68\pm0.34$    \\
nonzero  & 0         &$ 4.14\pm0.45$    &$3.93\pm0.46$   & $3.56\pm0.49$ &$4.51\pm0.45$       &$ 4.17\pm0.45$ &  $3.60\pm0.49$    \\
nonzero  & nonzero   & $4.04\pm0.46$   &$3.73\pm0.46$    &$3.26\pm0.45$  &$4.45\pm0.46$      &$4.00\pm0.46$  &   $3.31\pm0.45$    \\
  \hline 
 \end{tabular}
 \label{sphwmap}
\end{table*}
\end{document}